\documentclass[aps,prl,groupedaddress,showpacs,twocolumn,nofootinbib]{revtex4-2}
\usepackage{amsthm,amsmath,amssymb}
\usepackage{mathrsfs}
\usepackage{graphicx}
\usepackage{dcolumn}
\usepackage{bm}
\usepackage{amsfonts}
\usepackage{subfigure}
\usepackage{amsmath}
\usepackage[colorlinks=true,
linkcolor=blue,
citecolor=blue,
urlcolor=blue]{hyperref}
\usepackage{bbm}
\usepackage{braket}

\begin{document}
\title{Universal Error-Probability Scaling in Tolerance-Window Qualification of Quantum States}
\title{Universal Scaling of the Minimum Error Probability in Qualification of Quantum States}

\author{Zhaoyu Fei}
\affiliation{Zhejiang Key Laboratory of Quantum State Control and Optical Field Manipulation, Department of Physics, Zhejiang Sci-Tech University, 310018 Hangzhou, China}
\author{Yaotian Li}
\affiliation{Zhejiang Key Laboratory of Quantum State Control and Optical Field Manipulation, Department of Physics, Zhejiang Sci-Tech University, 310018 Hangzhou, China}
\author{Weicheng Huang}
\affiliation{Zhejiang Key Laboratory of Quantum State Control and Optical Field Manipulation, Department of Physics, Zhejiang Sci-Tech University, 310018 Hangzhou, China}
\author{Xiaoguang Wang}
\email{xgwang@zstu.edu.cn}
\affiliation{Zhejiang Key Laboratory of Quantum State Control and Optical Field Manipulation, Department of Physics, Zhejiang Sci-Tech University, 310018 Hangzhou, China}
\author{Y.-M. Du}
\email{ymdu@gscaep.ac.cn}
\affiliation{Graduate School of China Academy of Engineering Physics, Beijing 100193, China}

\date{\today}
\begin{abstract}
	Qualification of quantum states judges which of two sets of quantum states an unknown state lies in, where the two sets are labeled by two distinct parameter regions. We formulate this problem as a composite quantum hypothesis test and uncover universal scaling laws for the minimum error probability for $N$ copies. Taking polarization-direction qualification and purity qualification as examples, we show that the $N$-copy permutation symmetry and the geometric symmetries of the parameter regions identify the optimal measurements and the "worst pairwise states". The minimum error probability scales as $N^{-3/2}\exp(-N\xi)$ for disjoint regions and as $(NF)^{-1/2}$ for adjacent regions, where $\xi$ and $F$ are the quantum Chernoff divergence and quantum Fisher information associated with the "worst pairwise states", respectively. With the minimum error probability serving as an order parameter, the transition between the scaling behaviors becomes a second-order phase transition as $N\to\infty$.	Our approach determines whether a quantum state belongs to a given set without full state tomography, thereby enabling qualification of large ensembles using finite samples.

\end{abstract}

\maketitle

\textit{Introduction.}---
As quantum technologies progress from proof-of-principle demonstrations to practical implementations, a critical task is to judge whether experimentally prepared ensembles meet prescribed requirements, using only finite measurement resources. In this context, many quantum technologies do not require a source to produce a unique ideal state. Instead, its output parameters need to lie within a prescribed tolerance window~\cite{meyer2025quantum}, defined as the set of parameter values deemed acceptable for the intended operation. Such tolerance windows arise naturally in quantum technologies. Single-photon sources are assessed using thresholds on multiphoton suppression, indistinguishability, brightness, and mode quality~\cite{Santori2002,Somaschi2016,Ding2016,Lodahl2015,Senellart2017}. Quantum sensors and clocks require squeezing or metrological gain above classical benchmarks~\cite{Caves1981,Wineland1992,Kitagawa1993,Giovannetti2004,Giovannetti2011,Degen2017,Pezze2018,Schnabel2017}. Finite-sample quantum metrology asks whether a parameter encoded in a quantum state lies within a prescribed tolerance window~\cite{meyer2025quantum}. Quantum communication requires polarized photons transmitted through optical fibers to remain within an admissible angular region despite fiber-induced polarization diffusion~\cite{Czegledi2016,PhysRevLett.82.4815,WaiMenyuk1994,WaiMenyuk1995,WaiMenyuk1996,ImaiMatsumoto1987}.

While these platforms employ different acceptance criteria, a unified statistical framework for judging whether an unknown state lies within the corresponding admissible parameter region is still unestablished. Such a framework is needed to account for finite-sample fluctuations, calibration drift, device imperfections, and to assess device reliability, including aging and operational lifetime~\cite{PRL131160203,PRA112032220,Yi2026CapacityTime}.

In this Letter, we formulate the problem of qualifying an unknown quantum state according to prescribed acceptance criteria as a quantum composite hypothesis test. Two representative examples are considered: polarization-direction qualification (PDQ) and purity qualification (PQ). For both examples, the derivation exploits the permutation symmetry of the $N$ copies and the geometric symmetries of the parameter regions, namely $U(1)$ symmetry for PDQ and $SO(3)$ symmetry for PQ, to identify the optimal measurements and the worst pairwise states. These states minimize the quantum Chernoff divergence over two state sets~\cite{mosonyi2021error}. The relative configuration of the windows then gives rise to three universal finite-$N$ scaling laws for the minimum error probability $P_{\rm e}^{\rm min}$. For disjoint regions, $P_{\rm e}^{\rm min}\propto N^{-3/2}\exp(-N\xi)$; for adjacent regions, $P_{\rm e}^{\rm min}\propto(NF)^{-1/2}$; and for overlapping regions, $P_{\rm e}^{\rm min}$ converges to a nonzero intrinsic value as $N\to\infty$. Here, $\xi$ and $F$ are, respectively, the quantum Chernoff divergence~\cite{audenaert2007discriminating} and quantum Fisher information~\cite{helstrom1969quantum,Holevo1982,hubner1993computation,ma2009fisher,Liu_2020} associated with the worst pairwise states. Furthermore, in the $N\to\infty$ limit, $P_{\rm e}^{\rm min}$ serves as the order parameter for a second-order transition between the disjoint- and overlapping-region phases, with the adjacent-region configuration marking the critical point. Unlike quantum state tomography, which reconstructs the full state~\cite{James2001,Altepeter2005,Gross2010,Cramer2010,Flammia2011,DaSilva2011,Pallister2018,Eisert2020}, qualification of quantum states extracts only the relevant information and can therefore require fewer samples.

\begin{figure*}[htbp]
	\renewcommand{\figurename}{Fig.}
	\centering
	\includegraphics[scale=0.3]{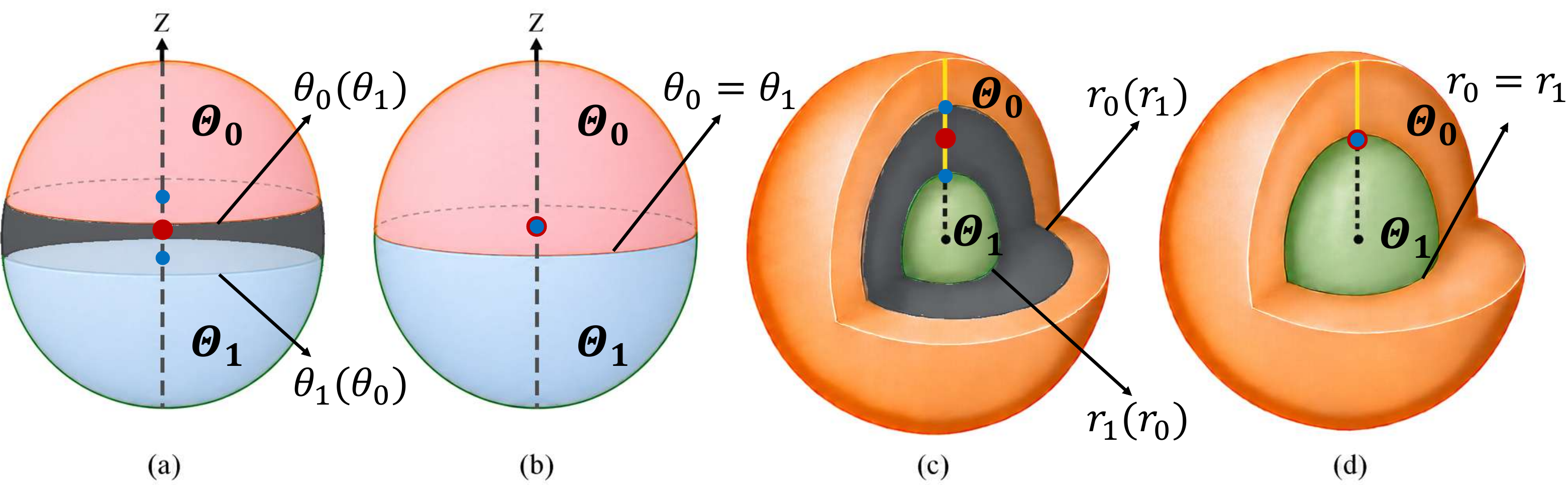}
	\caption{Schematic illustration of the region configurations for PDQ and PQ. Panels (a) and (b) show PDQ, whereas panels (c) and (d) show PQ. The gray domains in panels (a) and (c) have different meanings. For disjoint regions, they denote domains in which no state under consideration lies (for overlapping regions, they denote the intersections $\Theta_0\cap\Theta_1$). Panels (b) and (d) show adjacent regions. In panels (a) and (b), the orange and green surface domains denote $\Theta_0$ and $\Theta_1$, while the pink and light-blue interior domains denote the convex hulls of the corresponding state sets. In panels (c) and (d), the orange and green domains denote $\Theta_0$ and $\Theta_1$, respectively. In panel (c), the yellow line marks an arbitrary radius of the Bloch ball. In all panels, blue dots mark the worst pairwise states, and the red dot marks the cutoff state.}
	\label{total}
\end{figure*}

\textit{Qualification of quantum states.}---
Consider a quantum source with an output state $\rho_{\boldsymbol{\theta}}$, where $\boldsymbol{\theta}\in\Theta$. Qualification of quantum states partitions the parameter space $\Theta$ into an admissible region $\Theta_0$ and a rejected region $\Theta_1$, thereby defining two hypotheses,
$
H_0:\boldsymbol{\theta}\in\Theta_0,\
H_1:\boldsymbol{\theta}\in\Theta_1 .
$
These hypotheses are tested using a binary positive-operator-valued measure (POVM) $\{E_0,E_1\}$, with $E_0+E_1=\mathbbm{1}$ and $E_i\geq0$, where outcome $i$ judges that $H_i$ holds~\cite{audenaert2007discriminating}.

Because states associated with $\Theta_0$ are generally nonorthogonal to those associated with $\Theta_1$, perfect discrimination is impossible. A type-I error occurs when $\boldsymbol{\theta}\in\Theta_0$ is assigned to $H_1$, whereas a type-II error occurs when $\boldsymbol{\theta}\in\Theta_1$ is assigned to $H_0$. To reduce these errors, we use $N$ independent and identically prepared copies and optimize the POVM. Specifically, for a prior distribution ${\rm d}\mu(\boldsymbol{\theta})$ supported on $\Theta$, the Bayesian error probability is 
\begin{equation}
	P_{\rm e}
	=
	\sum_{i=0}^{1}
	\int_{\Theta_i}
	\operatorname{Tr}
	\left[
	\rho_{\boldsymbol{\theta}}^{\otimes N}
	E_{1-i}
	\right]
	{\rm d}\mu(\boldsymbol{\theta}) .
	\label{eq1}
\end{equation}
The optimal measurement is obtained by minimizing $P_{\rm e}$ over all binary POVMs. Introducing
$
	\rho_i
	=
	\pi_i^{-1}
	\int_{\Theta_i}
	\rho_{\boldsymbol{\theta}}^{\otimes N}
	{\rm d}\mu(\boldsymbol{\theta}),
$
with
$
	\pi_i
	=
	\int_{\Theta_i}
	{\rm d}\mu(\boldsymbol{\theta}),
$
the task reduces to binary state discrimination between $\rho_0$ and $\rho_1$ with priors $\pi_0$ and $\pi_1$. This optimization gives the minimum error probability~\cite{audenaert2007discriminating,Barnett:09,helstrom1969quantum}
\begin{equation}
	P_{\rm e}^{\rm min}
	=
	\frac{1}{2}
	\left[
	1
	-
	\left\|
	\pi_1\rho_1-\pi_0\rho_0
	\right\|_1
	\right],
	\label{eq:helstrom_min_error}
\end{equation}
where $\|\cdot\|_1$ denotes the trace norm. The optimal POVM is obtained according to the eigenspaces of $\pi_1\rho_1-\pi_0\rho_0$: $E_1$ projects onto its positive eigenspace and $E_0$ onto its negative eigenspace. The zero eigenspace may be arbitrarily assigned to $E_0$ or $E_1$.

Since the effective states $\rho_0$ and $\rho_1$ are defined by integrals of $\rho_{\boldsymbol{\theta}}^{\otimes N}$ over parameter regions, deriving the spectral decomposition of $\pi_1\rho_1-\pi_0\rho_0$ and the corresponding optimal POVM in closed form is generally difficult. We therefore exploit the symmetries shared by the regions and the prior distribution: these symmetries fix the structure of the optimal measurement and enable an analytic evaluation of the minimum error probability. We next illustrate this construction with two examples.

\begin{figure*}[htbp]
	\renewcommand{\figurename}{Fig.}
	\centering
	
	\includegraphics[scale=0.5]{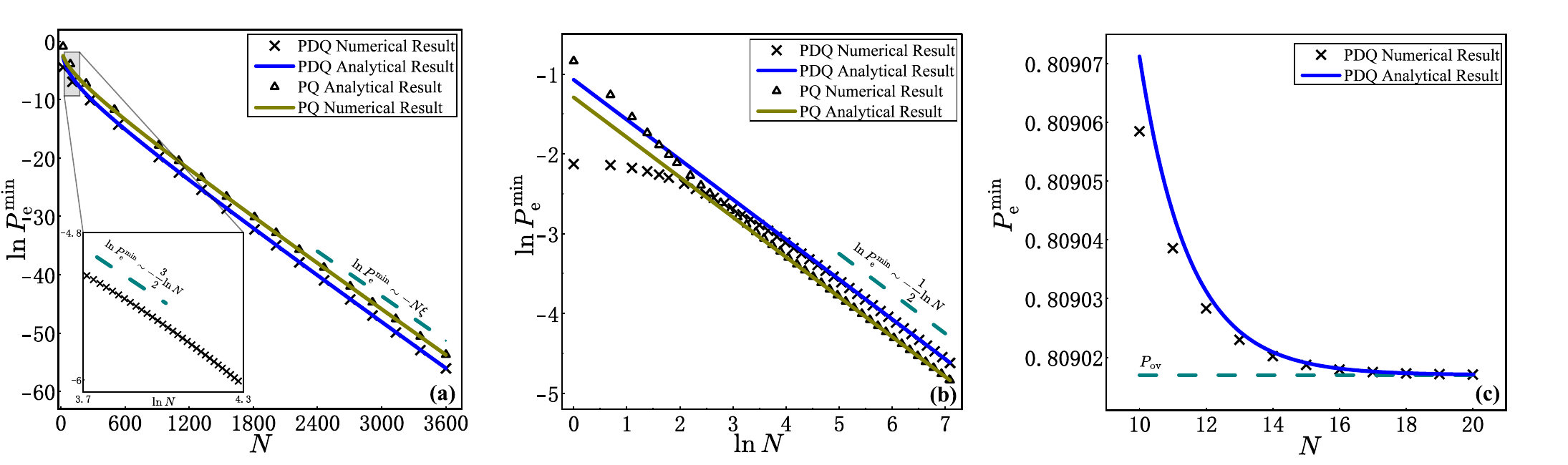}
	\caption{ Numerical verification of the minimum error probability for PDQ and PQ.
		(a) Disjoint regions, with $\theta_0=0.05\pi$, $\theta_1=0.15\pi$ for PDQ and $r_0=0.9$, $r_1=0.72$ for PQ.
		(b) Adjacent regions, with $\theta_0=\theta_1=0.15\pi$ for PDQ and $r_0=r_1=0.35$ for PQ.
		(c) Overlapping regions for PDQ, with $\theta_0=0.8\pi$ and $\theta_1=0.2\pi$.
		Crosses ($\times$) and triangles ($\triangle$) denote the numerical results for PDQ and PQ, respectively, obtained by  Eq.~\eqref{eq1}, whereas solid curves show the analytical results from Eq.~\eqref{p1}. Gray dashed lines indicate the asymptotic scaling laws in panels (a) and (b) and the intrinsic error probability in panel (c). The inset in panel (a) highlights the \(N^{-3/2}\) scaling in the small-\(N\) regime.
	}
	\label{perror}
\end{figure*}

\textit{Polarization-direction qualification and purity qualification.}---
We first consider the PDQ for a spin-$1/2$ state $|\psi\rangle$. The spin-$1/2$ state is parametrized by the Bloch-sphere angles $\boldsymbol{\theta}=(\theta,\phi)$ as
$
|\psi\rangle
=
\cos\frac{\theta}{2}\ket{1/2}
+
{\rm e}^{i\phi}\sin\frac{\theta}{2}\ket{-1/2}
$, where $\ket{1/2}$ and $\ket{-1/2}$ denote the states polarized along the $+z$ and $-z$ directions, respectively.
Choosing $+z$ as the target direction, the qualification is to judge whether the unknown direction lies within the prescribed angular regions, i.e.,
\begin{equation}
	\begin{split}
		\Theta_{0}^{\rm PDQ}
		&=
		\{(\theta,\phi):0\leq\theta\leq\theta_0,\;0\leq\phi<2\pi\},\\
		\Theta_{1}^{\rm PDQ}
		&=
		\{(\theta,\phi):\theta_1\leq\theta\leq\pi,\;0\leq\phi<2\pi\},\label{polarization}
	\end{split}
\end{equation}
where the boundaries $\theta_0$ and $\theta_1$ define polar caps that are invariant under rotations about the $z$ axis (the corresponding window configurations are illustrated in Fig.~\ref{total}a and~\ref{total}b). The prior distribution is chosen to share this $U(1)$ symmetry, i.e.,
$
{\rm d}\mu(\theta,\phi)
=
q\left(\cos\theta\right)
\sin\theta \mathrm{d}\theta \mathrm{d}\phi$ for some smooth density function $q$.

For PDQ, the permutation symmetry among the $N$ copies and the $U(1)$ symmetry of the parameter regions jointly determine the optimal measurement. Specifically, the permutation symmetry restricts the effective states $\rho_0$ and $\rho_1$ to the totally symmetric subspace of the $N$-copy Hilbert space $\mathcal{H}_{1/2}^{\otimes N}$, with total spin number $S_{\max}=N/2$. Moreover, the $U(1)$ symmetry further renders $\pi_1\rho_1-\pi_0\rho_0$ diagonal in the Dicke basis $\{\ket{S_{\max},M}\}$, where $M$ is the magnetic quantum number. The remaining subspaces can be arbitrarily assigned to $E_0$ or $E_1$, since $\pi_1\rho_1-\pi_0\rho_0$ has zero eigenvalues on them. This freedom reduces the Dicke-basis measurement to an equivalent local implementation: each spin is measured along the $z$ direction, and the judgment is made from the total magnetization. 


Guided by the symmetry analysis above, we parametrize the POVM by cutoff angle $x$~\cite{cutoff}  as
\begin{equation}
	\begin{split}
		E_0^{\rm PDQ}(x)
		&=
		\sum_{M=S_{\rm max}\cos x+1}^{S_{\rm max}}
		|\boldsymbol{S}\rangle_{M} {}_{M}\langle\boldsymbol{S}| ,
	\end{split}
	\label{E0p}
\end{equation}
where
$
|\boldsymbol{S} \rangle_M
=
|s_1\rangle\otimes\cdots\otimes|s_N\rangle
$
denotes a product state in the $z$ basis, with $s_k\in\{1/2,-1/2\}$ constrained by
$
M
=
\sum_{k=1}^{N}s_k
$. 
Substituting Eq.~\eqref{E0p} into Eq.~\eqref{eq1} gives the $x$-dependent error probability $P_{\rm e}(x)$. The minimizer $x_*$ is then obtained from ${\rm d} P_{\rm e}(x)/{\rm d}x=0$ (see Supplemental Material ~\cite{supplement} for details), yielding
\begin{equation}
	x_* =
	\begin{cases}
		\arccos\left[ \dfrac{
			\ln\left( \sin\theta_1\right)
			-
			\ln\left(\sin\theta_0\right)
		}{
			\ln\left( \tan\frac{\theta_1}{2}\right)
			-
			\ln\left(\tan\frac{\theta_0}{2} \right)
		}\right] ,
		& \theta_0 \neq \theta_1, \\[8pt]
		\theta_0,
		& \theta_0=\theta_1 .\label{x1}
	\end{cases}
\end{equation}
Substituting Eq.~\eqref{x1} into $P_{\rm e}(x)$ gives
\begin{equation}
	P_{\rm e}^{\rm min}
	\simeq
	\left\{
	\begin{array}{ll}
		A_{\rm s}
		N^{-3/2}\mathrm{e}^{-N\xi},
		& \theta_0<\theta_1,
		\\[2mm]
		A_{\rm b}
		\left[NF\right]^{-1/2},
		& \theta_0=\theta_1,
		\\[2mm]
		P_{\rm ov}
		+
		A_{\rm s}
		N^{-3/2}\mathrm{e}^{-N\xi},
		& \theta_0>\theta_1.
	\end{array}
	\right.
	\label{p1}
\end{equation}
where $A_{\rm s}$ and $A_{\rm b}$ are $N$-independent
prefactors given in  Supplemental Material ~\cite{supplement} (when the prior density vanishes at the boundary, the scaling laws of minimum error probability are modified~\cite{notePriorBoundary}).
The three lines correspond to disjoint, adjacent, and overlapping
regions, respectively. For disjoint regions, the exponential decay is characterized by
the Chernoff exponent $\xi$~\cite{mosonyi2021error}, which is given by
\begin{equation}
	\xi
	=
	D(t_*\|t_0)
	=
	D(t_*\|t_1)
	=
	C(t_0\|t_1),
	\label{classcher}
\end{equation}
where
$
t_i=\cos^2(\theta_i/2)
$,
$
t_*=\cos^2(x_*/2)
$,
$
D(a\|b)
=
a\ln(a/b)
+
(1-a)\ln[(1-a)/(1-b)]
$
and
$
C(a\|b)
=
-\ln\inf_{0<s<1}
\left[
a^s b^{1-s}
+
(1-a)^s(1-b)^{1-s}
\right]
$
are respectively the Kullback--Leibler (KL) divergence and the Chernoff divergence between two Bernoulli
distributions~\cite{chernoff1952measure,
	hellman1970probability,blahut1974hypothesis}. In the third equality of Eq.~\eqref{classcher}, we have used the relation between the KL divergence and the Chernoff divergence~\cite{nielsen2022revisiting}. For adjacent regions, the exponential decay is
replaced by a power law. As $t_1\to t_0$,
$
C(t_0\|t_1)
\sim
F(t_0-t_1)^2/8,
$
where
$
F=1/[t_0(1-t_0)]
$
is the Fisher information of the Bernoulli distribution entering Eq.~\eqref{p1}.
For overlapping regions, the intersection $\Theta_0\cap\Theta_1$ carries the prior weight~\cite{notePriorWeight} and
$
P_{\rm ov}
=
\pi_0+\pi_1-1,
$
which gives a nonzero intrinsic error probability as $N\to\infty$. The Chernoff exponent $\xi$ also governs the finite-$N$ correction to this limit.

We next consider the PQ  for a qubit state $\rho$. The qubit state is parametrized by the Bloch-vector $\mathbf r=(r_x, r_y, r_z)$ as
$
\rho
=
\left(
\mathbbm{1}
+
\mathbf r\cdot\boldsymbol{\sigma}
\right)/2,
$
where $0\le r\le 1$ and $\boldsymbol{\sigma}=(\sigma_x,\sigma_y,\sigma_z)$ is the vector of Pauli matrices.
Since the purity of $\rho$ is $\operatorname{Tr}(\rho^2)=(1+r^2)/2$, the qualification is to judge whether the Bloch radius $r$ lies within the prescribed radial tolerance in spherical coordinates on the Bloch ball, i.e.,
\begin{equation}
	\begin{split}
		\Theta_0^{\rm PQ}
		&=
		\{(r,\theta,\phi):r_0\leq r\leq 1,\;0\leq\theta\leq\pi,\;0\leq\phi<2\pi\},\\
		\Theta_1^{\rm PQ}
		&=
		\{(r,\theta,\phi):0\leq r\leq r_1,\;0\leq\theta\leq\pi,\;0\leq\phi<2\pi\},\label{purity}
	\end{split}
\end{equation}
where $r_0$ and $r_1$ denote the radial boundaries. The regions defined by these boundaries are invariant under arbitrary rotations of the Bloch vector (the corresponding windows configurations are illustrated in Fig.~\ref{total}c and \ref{total}d). The prior distributions are chosen to share this $SO(3)$ symmetry, i.e.,
${\rm d}\mu(r,\theta,\phi)=w(r)r^2\sin\theta\,{\rm d}r\,{\rm d}\theta {\rm d}\phi$, for some  smooth density function $w$.

For the PQ, the $N$-copy state remains permutation invariant, but the $U(1)$ symmetry about the $z$-axis is enlarged to $SO(3)$ rotational symmetry about the Bloch vector. As in the PDQ, these two symmetries determine the structure of $\pi_1\rho_1-\pi_0\rho_0$: it is constant on each block labeled by the total angular momentum $S$, with the constant $\lambda_S$. More explicitly, Schur--Weyl duality gives the decomposition
$ \mathcal{H}_{1/2}^{\otimes N} = \bigoplus_S \mathcal{V}_S\otimes\mathcal{K}_S$, which together with Schur's lemma leads to $ \pi_1\rho_1-\pi_0\rho_0 = \bigoplus_S \lambda_S \left( \mathbbm{1}_{\mathcal{V}_S}\otimes \mathbbm{1}_{\mathcal{K}_S}\right) $~\cite{bartlett2007reference}. Here $\mathcal{V}_S$ carries the irreducible $SU(2)$ repres   entation with total angular momentum $S$, while $\mathcal{K}_S$ is the multiplicity space carrying the corresponding irreducible representation of the symmetric group $\mathcal{S}_N$. 

Guided by the symmetry analysis above, we parametrize the POVM by a cutoff $x$~\cite{cutoff} as
\begin{equation}
	\begin{split}
		E_0^{\rm PQ}(x)
		&=
		\bigoplus_{S=xS_{\rm max}+1}^{S_{\rm max}}  \left( \mathbbm{1}_{\mathcal{V}_S}\otimes \mathbbm{1}_{\mathcal{K}_S}\right).\label{E0pu}
	\end{split}
\end{equation}
Substituting Eq.~\eqref{E0pu} into Eq.~\eqref{eq1} gives the $x$-dependent error probability $P_{\rm e}(x)$. Applying a minimization procedure similar to that used for the PDQ yields the minimizer $x_*$ and the minimum error probability for all three types of region configurations. Their expressions take the same forms as in Eqs.~\eqref{x1} and \eqref{p1}, with $t_i$ and $t_*$ now redefined as $t_i=(1+r_i)/2$ and $t_*=(1+x_*)/2$.
The explicit prefactors are given in Supplemental Material ~\cite{supplement}.

These two examples exhibit universal scaling of the minimum error probability: despite differences in their prior distributions, symmetries, and optimal measurements, both exhibit the three types of scaling laws summarized in Eq.~\eqref{p1}.  In Fig.~\ref{perror}, we further compare the numerical results obtained by minimizing the error probability defined in Eq.~\eqref{eq1} with the analytical asymptotic expressions in Eq.~\eqref{p1} for both qualification tasks under the three types of region configurations. Their agreement at large $N$ supports the predicted universal scaling. The reference dashed lines indicate the $N^{-3/2}$ prefactor and the $\exp(-N\xi)$ decay in panel (a), the $N^{-1/2}$ scaling in panel (b), and the intrinsic error probability in panel (c).

\textit{Worst pairwise states and phase transition.}---The worst pairwise states, denoted by $\rho_{{\rm w},0}$ and $\rho_{{\rm w},1}$, are a pair of states that attain the minimum quantum Chernoff divergence over all pairs belonging to two sets of quantum states, $\rho\in\mathcal{C}_0$ and $\sigma\in\mathcal{C}_1$, where the quantum Chernoff divergence is defined as $
C_{\rm Q}(\rho\|\sigma)
=
-\ln
\min_{0\leq s\leq1}
\operatorname{Tr}\!\left(
\rho^s\sigma^{1-s}
\right)
$~\cite{fuchs1996distinguishability,audenaert2007discriminating,mosonyi2021error}.
 When the two worst pairwise states approach each other through a parameter, i.e.,
 $
 \rho_{{\rm w},0}=\rho_{\alpha}
 $
 and
 $
 \rho_{{\rm w},1}=\rho_{\alpha+{\rm d}\alpha},
 $
 the leading variation of the quantum Chernoff divergence is
 $
 	C_{\rm Q}
 	\left(
 	\rho_{\alpha}
 	\middle\|
 	\rho_{\alpha+{\rm d}\alpha}
 	\right)
 	\sim
 	F_{\rm Q}
 	\left(
 	\rho_{\alpha}
 	\right)
 	({\rm d}\alpha)^2/8
$~\cite{calsamiglia2008quantum},
 where $F_{\rm Q}(\rho_{\alpha})={\rm Tr}\left(\rho_{\alpha}L_{\alpha}^2\right) $ is the quantum Fisher information with respect to the parameter $\alpha$ and $L_{\alpha}$ is the symmetric logarithmic derivative defined by $\partial\rho_{\alpha}/\partial{\alpha}=\left( \rho_{\alpha}L_{\alpha}+L_{\alpha}\rho_{\alpha}\right)/2 $~\cite{nussbaum2009chernoff,braunstein1994statistical,Liu_2020}. 
 
 We next use the worst pairwise states to reveal the intrinsic connection between the geometric symmetry and the universal scaling of the minimum error probability. In the PDQ, the two sets of quantum states are the convex hulls~\cite{noteConvexHull} of the state families associated with the parameter regions in Eq.~\eqref{polarization},
 $
 \mathcal{C}_i
 =
 \operatorname{conv}
 \{
 \rho_{\boldsymbol{\theta}}:
 \boldsymbol{\theta}\in\Theta_i
 \}.
 $

 The $U(1)$ symmetry of $\mathcal{C}_0$ and $\mathcal{C}_1$ restricts the search for the worst pairwise states to the $z$ axis. Among the corresponding intersection points, the pair minimizing the quantum Chernoff divergence gives the worst pairwise states shown in Fig.~\ref{total}(a) and \ref{total}(b)  (see Supplemental Material ~\cite{supplement} for details). Their density matrices are diagonal in the $S_z$ basis,
 $
 \rho_{{\rm w},i}
 =
 \operatorname{diag}
 \left[
  t_i,
  1-t_i
 \right],
 $
  where $
  t_i=\cos^2(\theta_i/2)
  $. Their quantum Chernoff divergence therefore coincides with the Chernoff divergence between two Bernoulli distributions in Eq.~\eqref{classcher}. The relation derived above between the KL divergence and the Chernoff divergence has a quantum counterpart: 
$
 	C_{\rm Q}\left(
 	\rho_{{\rm w},0}
 	\middle\|
 	\rho_{{\rm w},1}
 	\right)
 	=
 	D_{\rm Q}
 	\left(
 	\rho_{x_*}
 	\middle\|
 	\rho_{{\rm w},0}
 	\right)
 	=
 	D_{\rm Q}
 	\left(
 	\rho_{x_*}
 	\middle\|
 	\rho_{{\rm w},1}
 	\right),
$
  where
 $
 D_{\rm Q}(\rho\|\sigma)
 =
 \operatorname{Tr}
 [
 \rho(\ln\rho-\ln\sigma)
 ]
 $
 is the quantum relative entropy~\cite{hiai1991proper}.  Here, the cutoff state (marked by the red dot in Fig.~\ref{total}a) is diagonal in the $S_z$ basis,
 $
 \rho_{x_*}
 =
 \operatorname{diag}
 [
 t_*,
 1-t_*
 ]
 $
 with $x_*$ given by Eq.~\eqref{x1}, where $
 t_*=\cos^2(x_*/2)
 $. For the adjacent-region configuration, the three states coincide,
 $
 \rho_{{\rm w},0}
 =
 \rho_{{\rm w},1}
 =
 \rho_{x_*},
 $
 as shown in Fig.~\ref{total}(b), of which the quantum Fisher information reproduces the Fisher information of the Bernoulli distribution in Eq.~\eqref{p1}. For the overlapping-region configuration, the minimum error probability approaches a nonzero intrinsic value.

In the PQ, the  two sets of quantum states are taken as the quantum states  associated with the parameter regions in Eq.~\eqref{purity}, $
\mathcal{C}_i
=
\{
\rho_{\boldsymbol{\theta}}:
\boldsymbol{\theta}\in\Theta_i
\}
$. The $SO(3)$ symmetry makes all radial directions of the Bloch ball equivalent. For the disjoint- and adjacent-region cases, the density matrices of worst pairwise states are the pair of intersections of any chosen radius with $\mathcal{C}_0$ and $\mathcal{C}_1$ that minimizes the quantum Chernoff divergence, as shown in Fig.~\ref{total}(c) and \ref{total}(d) (see Supplemental Material ~\cite{supplement} for details). In the basis aligned with the chosen radius, the worst pairwise states in the PQ have the same diagonal structure as those in the PDQ, except that $t_i$ and $t_*$ are now redefined as $t_i=(1+r_i)/2$ and $t_*=(1+r_*)/2$. The preceding analysis therefore yields analogous results for the Chernoff divergence and the Fisher information in Eqs.~\eqref{p1} and \eqref{classcher}.

\begin{figure}[t]
	\renewcommand{\figurename}{Fig.}
	\centering
	
	\includegraphics[scale=0.3]{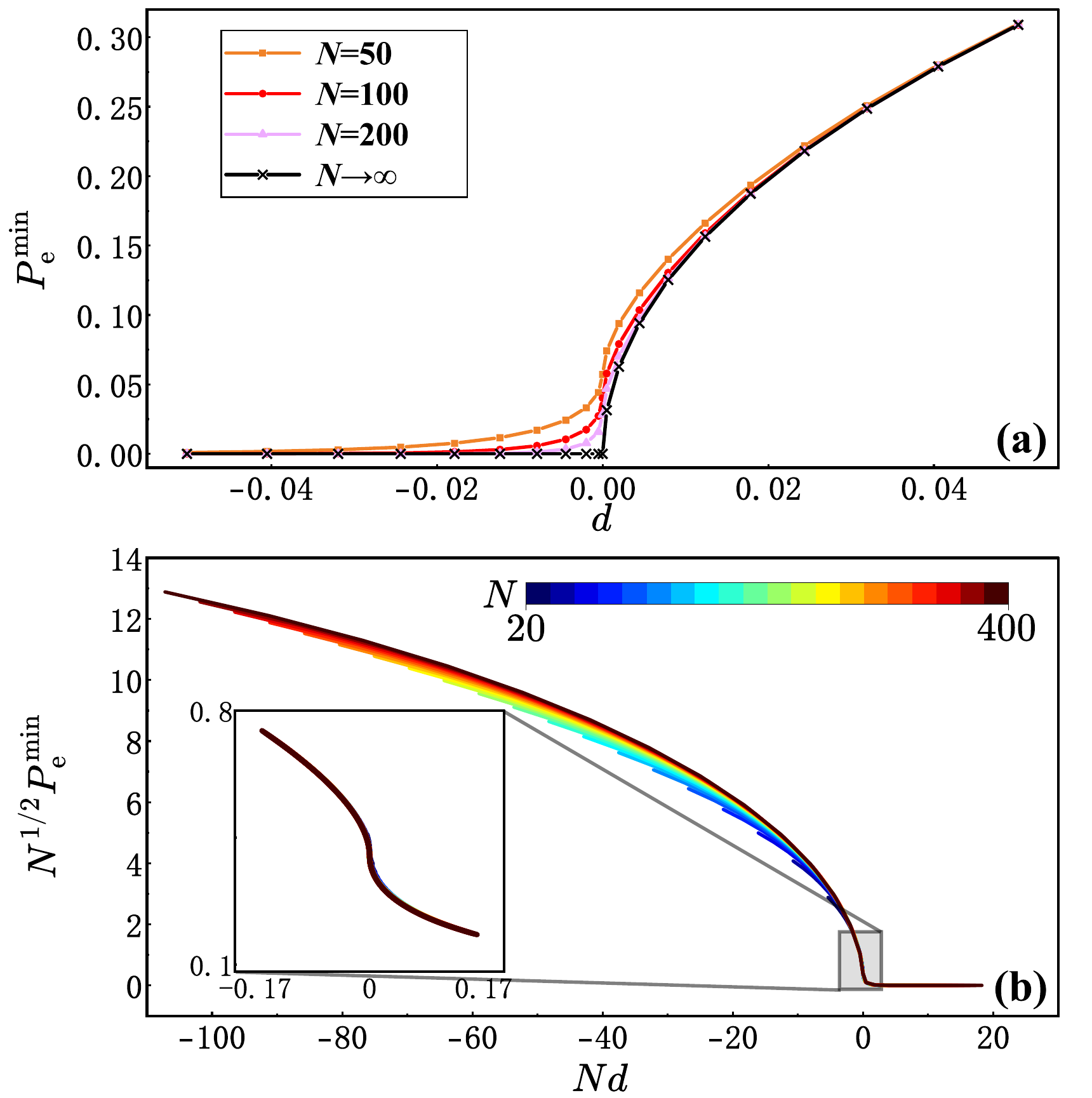}
	\caption{Finite-$N$ scaling analysis of the minimum error probability in the PDQ.
		(a) Minimum error probability $ P_{\rm e}^{\rm min}$ as a function of signed distance for $N$ copies.
		(b) For $N$ copies, with $N$ from $N=20$ to $N=400$, the rescaled error probability $N^{1/2} P_{\rm e}^{\rm min}$ is plotted as a function of $Nd$.
	}
	\label{phase}
\end{figure}

The three types of scaling laws in Eq.~\eqref{p1} reveal a phase transition characterized by the signed distance $d=\epsilon\xi$. Here, $\epsilon$ takes values $-1$, $0$, and $1$ for disjoint, adjacent, and overlapping regions, respectively.  The adjacent-region case then marks the critical point. For PDQ near the critical point,
$P_{\rm ov}/\sqrt{\xi}$, $\xi A_{\rm s}$, and the prefactor
$A_{\rm b}$ in Eq.~\eqref{p1} depend only on the cutoff angle
$x_*$ (see Supplemental Material~\cite{supplement} for details). To eliminate this nonuniversal dependence, we impose
$
\theta_0+\theta_1=\pi,
$
thereby fixing $x_*=\pi/2$.
 Figure~\ref{phase}(a) shows $P_{\rm e}^{\rm min}$ as a function of $d$ for several copy numbers $N$.  As $N$ increases, the transition from the disjoint-region phase ($d<0$) to the overlapping-region phase ($d>0$) becomes sharper, with $P_{\rm e}^{\rm min}$ as an order parameter that remains finite in the overlapping-region phase and vanishes in the disjoint-region phase. This behavior is consistent with a second-order transition in the large-$N$ limit.   In Fig.~\ref{phase}(b), we plot $N^{1/2}P_{\rm e}^{\rm min}$ as a function of $Nd$ for copy numbers from $N=20$ to $N=400$ and observe a collapse near the critical point, which is described by the finite-$N$ scaling form
\begin{equation}
	P_{\rm e}^{\rm min}(N,d,x_*)
	=
	N^{-1/2}
	g(x_*)f(x_*,Nd),
\end{equation}
 where $g(x_*)$ is a prefactor determined by the prior distribution, while $f(x_*,Nd)$ is a scaling function.

\textit{Discussions.}---Previous work in composite quantum hypothesis testing has reduced the Chernoff exponent governing the minimum error probability for discriminating two quantum-state sets to the Chernoff divergence between their worst pairwise states, typically assuming commuting or quasi-classical state families, or convex state sets~\cite{Berta2021,mosonyi2021error,Lami2025Stein,Lami2025Sanov,FangHayashi2025Chernoff,SimpsonPaliasJose2026}. The parameter regions considered here generally violate these assumptions: the corresponding sets of quantum states are nonconvex in the state space, and states drawn from the competing regions can be both noncommuting and nonorthogonal.  Nevertheless, by exploiting the permutation symmetry of the $N$ copies and the geometric symmetries of the parameter regions, we reduce the Chernoff exponent to the Chernoff divergence between the worst pairwise states and identify these states via the symmetries of the state sets. This reduction therefore extends beyond the previous commuting, quasi-classical, and convex settings.

To illustrate the application of the theory, consider photons with horizontal ($H$) or vertical ($V$) polarization are disturbed during the long-distance transmission in optical fiber. 
We aim to qualify photon polarization using PDQ.  Let $\gamma$ denote the ratio of the transmission distance to the polarization-diffusion length of the optical fiber~\cite{Galtarossa2000BeatLength,Galtarossa2001CorrelationLength,Eronyan2021PMFiber}, with $\gamma<\ln 2$, $\gamma=\ln 2$, and $\gamma>\ln 2$ corresponding to the disjoint-, adjacent-, and overlapping-region cases in PDQ, respectively. For $\gamma<\ln 2$, our result suggests that achieving an error probability $\epsilon$ requires a detected photon number $N_{\rm req}\simeq2\ln\epsilon /\ln\left[1-\left(2\exp(-\gamma)-1 \right)^2  \right] $, while $\gamma=\ln 2$ corresponds to the maximum fiber transmission distance over which photon polarizations remain distinguishable, beyond which an intrinsic error prevents the qualification error probability from being made arbitrarily small by increasing the photon number.

\textit{Acknowledgments.}---This work was supported by the Science Challenge Project (Grant No.~TZ2025017) , the Quantum Science and Technology-National Science and Technology Major Project (Grant No.~2024ZD 0301000) , the National Natural Science Foundation of China (Grant No.~12405046), and the Science Foundation of Zhejiang Sci-Tech University (Grants No.~23062088-Y and No.~23062181-Y).

\textit{Data availability.}---The data that support the findings of
this article are openly available~\cite{fei_2026_22040993}.

\bibliographystyle{apsrev4-2}
\bibliography{refs2}

\newpage
\begin{widetext}
	\section{Supplemental Material for
		“Universal Scaling of the Minimum Error Probability
		in Qualification of Quantum States”}
	\section{The Optimal Measurement for PDQ}
	
	In this part, we derive the optimal measurement for PDQ by substituting a projective measurement with a cutoff angle $x$ into the error probability. For disjoint regions ($\theta_0<\theta_1$), the projective measurement is defined as
	\begin{equation}
		\begin{split}
			E_0^{\rm PDQ}(x)
			&=
			\sum_{M=S_{\rm max}\cos x+1}^{S_{\rm max}}
			|\boldsymbol{S}\rangle_{M} {}_{M} \langle\boldsymbol{S}| .
		\end{split}
		\label{E1p}
	\end{equation}
	Substituting Eq.~\eqref{E1p} and $S_{\rm max}=\frac{N}{2}$ into the expression for the error probability gives
	\begin{equation}
		\begin{split}
			P_{\rm e}(x)
			&=
			4\pi\Bigg[
			\sum_{M=-\frac{N}{2}}^{\frac{N}{2}\cos x}
			\binom{N}{\frac{N}{2}+M}
			\int_{t_0}^{1}
			q\left(2t-1\right)t^{M+\frac{N}{2}}(1-t)^{\frac{N}{2}-M}{\rm d}t \\
			&+
			\sum_{M=\frac{N}{2}\cos x+1}^{N/2}
			\binom{N}{\frac{N}{2}+M}
			\int_{0}^{t_1}
			q\left(2t-1\right)t^{\frac{N}{2}+M}(1-t)^{\frac{N}{2}-M}{\rm d}t
			\Bigg]\label{error}.	\end{split}
	\end{equation}
	Here, we set $t=\cos^2(\theta/2)$ and
	$t_i=\cos^2(\theta_i/2)$, with $i=0,1$. The factor
	$\binom{N}{\frac{N}{2}+M}$ is the binomial coefficient, defined by
	$
	\binom{N}{\frac{N}{2}+M}
	=
	N!/
	[\left(\frac{N}{2}-M\right)!
	\left(\frac{N}{2}+M\right)!]
	$. Accordingly, the first term is the contribution from states in the admissible region that are assigned to $H_1$, while the second term is the contribution from states in the rejected region that are assigned to $H_0$. The binomial coefficients in the error probability originate from
	the expansion of the $N$-copy state in the $|\boldsymbol{S}\rangle_{M}$
	basis:~\cite{PhysRevA.101.023601}
	\begin{equation}
		(\ket{\psi(\theta,\phi)})^{\otimes N}
		=
		\sum_{M=-N/2}^{N/2}
		\sqrt{\binom{N}{\frac{N}{2}+M}}
		\left( \cos\frac{\theta}{2}\right) ^{\left(\frac{N}{2}+M\right)}
		\left( \sin\frac{\theta}{2}\right) ^{\left(\frac{N}{2}-M\right)}
		{\rm e}^{ i\left(\frac{N}{2}-M\right)\phi}
		|\boldsymbol{S}\rangle_{M} .
	\end{equation} 
	
	For large $N$, the cutoff angle $x$ can be treated as a continuous variable. The stationarity condition $\left.{\rm d}P_{\rm e}(x)/{\rm d}x\right|_{x=x_*}=0$ is, at leading order, equivalent to the equality of the two boundary contributions at the cutoff angle $x=x_*$:
	\begin{equation}
		\int_{t_0}^{1}
		q\left(2t-1\right)t^{t_*N}(1-t)^{(1-t_*)N}{\rm d}t
		=
		\int_{0}^{t_1}
		q\left(2t-1\right)t^{t_*N}(1-t)^{(1-t_*)N}{\rm d}t ,
	\end{equation}
	where $t_*=\cos^2(x_*/2)$. In the large-$N$ limit, the two integrals are dominated by
	their respective boundary points, $t_0$ and $t_1$. Therefore, upon retaining only the leading exponential contributions, the stationarity condition reduces to
	\begin{equation}
		t_0^{t_*N}(1-t_0)^{(1-t_*)N}
		=
		t_1^{t_*N}(1-t_1)^{(1-t_*)N} .
	\end{equation}
	Taking the logarithm and solving for $x_*$ yields
	\begin{equation}
		x_*=	\arccos\left[ \dfrac{
			\ln\left( \sin\theta_1\right)
			-
			\ln\left(\sin\theta_0\right)
		}{
			\ln\left( \tan\frac{\theta_1}{2}\right)
			-
			\ln\left(\tan\frac{\theta_0}{2} \right)
		}\right].\label{sex_*}
	\end{equation}
	For the adjacent-region case
	($\theta_0=\theta_1$), the corresponding cutoff is obtained from
	the disjoint-region result by taking the limit
	$\theta_1\to\theta_0$. Applying this limit to
	Eq.~\eqref{sex_*} yields
	\begin{equation}
		x_*
		=
		\lim_{\theta_1\to \theta_0}\arccos\left[ \dfrac{
			\ln\left( \sin\theta_1\right)
			-
			\ln\left(\sin\theta_0\right)
		}{
			\ln\left( \tan\frac{\theta_1}{2}\right)
			-
			\ln\left(\tan\frac{\theta_0}{2} \right)
		}\right] = \theta_0.
	\end{equation}
	For the overlapping-region case
	($\theta_0>\theta_1$), the cutoff $x_*$ is given by the same
	expression as in the disjoint-region case.
	
	\section{The Minimum Error Probability for PDQ}

	In this part, we derive the asymptotic form of the minimum error
	probability for PDQ. We use the relation between the regularized
	incomplete beta function and the binomial cumulative distribution:
	\begin{equation}
		I_{1-p}(n-k,k+1)
		=
		\sum_{j=0}^{k}
		\binom{n}{j}
		p^j(1-p)^{n-j},\label{betabino}
	\end{equation}
	where $I_{1-p}(n-k,k+1)=\int_0^{1-p}t^{n-k-1}(1-t)^k{\rm d}t/\int_0^{1}t^{n-k-1}(1-t)^k{\rm d}t$, as defined in Ref.~\cite{Olver2010NIST}. Substituting Eqs.~\eqref{betabino}
	and~\eqref{sex_*} into Eq.~\eqref{error}, we obtain the minimum
	error probability in the following form:
	\begin{equation}
		\begin{split}
			P_{\rm e}^{\rm min}
			=&4\pi\Bigg[\frac{\int_{t_0}^1{\rm d}t't'^{t_*N}(1-t')^{(1-t_*)N-1}\int_{t_0}^{t'}{\rm d}tq(2t-1)}{\int_{0}^{1}{\rm d}t't'^{t_*N}(1-t')^{(1-t_*)N-1}}+\frac{\int_{0}^{t_1}{\rm d}t't'^{t_*N}(1-t')^{(1-t_*)N-1}\int_{t'}^{t_1}{\rm d}tq(2t-1)}{\int_{0}^{1}{\rm d}t't'^{t_*N}(1-t')^{(1-t_*)N-1}}\Bigg]\\
			=&4\pi\Bigg[\left[ \mathcal Q(t_*)-\mathcal Q(t_0)\right] I_{1-t_0}\left((1-t_*)N, t_*N+1\right)+\frac{\int_{t_0}^1{\rm d}t'\mathrm{e}^{-(N-1)D(y\|t') } \left[\mathcal Q(t')-\mathcal Q(t_*)\right]}{\int_{0}^{1}{\rm d}t'\mathrm{e}^{-(N-1)D(y\|t')}}\\
			&+\left[ \mathcal Q(t_1)-\mathcal Q(t_*)\right] I_{t_1}\left(t_*N+1, (1-t_*)N\right)+\frac{\int_{0}^{t_1}{\rm d}t'\mathrm{e}^{-(N-1)D(y\|t') }\left[\mathcal Q(t_*)-\mathcal Q(t')\right]}{\int_{0}^{1}{\rm d}t'\mathrm{e}^{-(N-1)D(y\|t')  }}\Bigg],\label{13}
		\end{split}
	\end{equation}
	where $y\equiv Nt_*/(N-1)$, $D(a\|b)=a\ln(a/b)+(1-a)\ln[(1-a)/(1-b)]$ denotes the Kullback--Leibler divergence and
	$
	\mathcal Q(b)$ is defined as $
	\mathcal Q(b)=\int_0^b q(2a-1){\rm d}a 
	$. Applying integration by parts to the two integrals appearing in the numerators and expanding the resulting expressions for large $N$,
	we obtain
	\begin{equation}
		\begin{split}
			P_{\rm e}^{\rm min}
			\simeq	&4\pi\Bigg\{ [Q(t_0)-Q(t_*)]\times\\
			&\left[-I_{1-t_0}\left((1-t_*)N, t_*N+1\right)+\frac{\mathrm{e}^{-(N-1)D(y\|t_0)}\sqrt{(N-1)F}}{\sqrt{2\pi}(N-1)\partial_{t_0}D(y\|t_0)}-\frac{\partial_{t_0}^2D(y\|t_0)\mathrm{e}^{-(N-1)D(y\|t_0)}\sqrt{(N-1)F}}{\sqrt{2\pi}(N-1)^2\left(\partial_{t_0}D(y\|t_0)\right)^3 } \right]\\
			&+[Q(t_1)-Q(t_*)]\times\\
			&\left[I_{t_1}\left(t_*N+1,(1-t_*)N\right)+\frac{\mathrm{e}^{-(N-1)D(y\|t_1)}\sqrt{(N-1)F}}{\sqrt{2\pi}(N-1)\partial_{t_1}D(y\|t_1)}-\frac{\partial_{t_1}^2D(y\|t_1)\mathrm{e}^{-(N-1)D(y\|t_1)}\sqrt{(N-1)F}}{\sqrt{2\pi} (N-1)^2\left(\partial_{t_1}D(y\|t_1)\right)^3} \right]\\
			& +\frac{q(2t_0-1)\mathrm{e}^{-(N-1)D(y\|t_0)}\sqrt{(N-1)F}}{\sqrt{2\pi}(N-1)^2\left(\partial_{t_0}D(y\|t_0) \right)^2}+\frac{q(2t_1-1)\mathrm{e}^{-(N-1)D(y\|t_1)}\sqrt{(N-1)F}}{\sqrt{2\pi}(N-1)^2\left(\partial_{t_1}D(y\|t_1) \right)^2}\Bigg\},\label{15}
		\end{split}
	\end{equation}
	where $\int_0^1\exp[-(N-1)D(y\|t)]{\rm d}t\simeq\sqrt{2\pi/[(N-1)F(y)]}$ and $F=1/(t_*(1-t_*))$ is the Fisher information of the Bernoulli distribution.

	\begin{figure*}[htbp]
		\renewcommand{\figurename}{Fig.}
		\centering
		
		\includegraphics[scale=0.6]{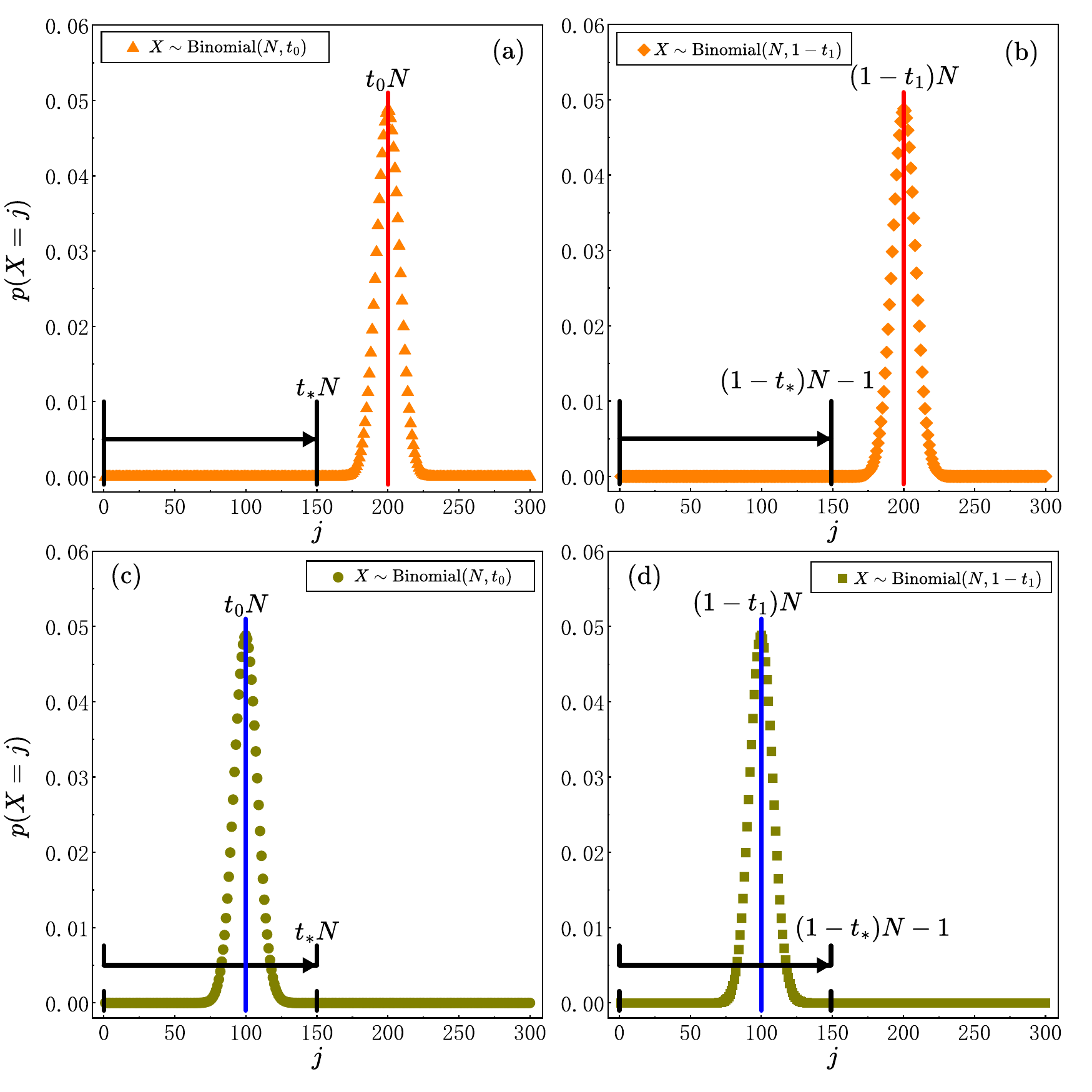}
		\caption{The points in panels (a) and (c) show the probability mass functions of $X\sim{\rm Binomial}(N,t_0)$ , while those in panels (b) and (d) show the probability mass functions of $X\sim{\rm Binomial}(N,1-t_1)$. In panels (a) and (b), the parameters are $t_0=2/3$, $t_1=1/3$, $t_*=1/2$, and $N=300$. In panels (a) and (b), the parameters are $t_0=1/3$, $t_1=2/3$, $t_*=1/2$, and $N=300$. The black arrows in panels (a)--(d) indicate the corresponding summation ranges.}
		\label{probability}
	\end{figure*}
	
	Using the beta--binomial relation in Eq.~\eqref{betabino}, the two regularized incomplete
	beta functions relevant to the disjoint-region asymptotics can be
	written as:
	\begin{equation}
		\begin{aligned}
			I_{1-t_0}\!\left((1-t_*)N,t_*N+1\right)
			=
			\sum_{j=0}^{t_*N}
			\binom{N}{j}
			t_0^j(1-t_0)^{N-j},\quad
			I_{t_1}\!\left(t_*N+1,(1-t_*)N\right)
			=
			\sum_{j=0}^{(1-t_*)N-1}
			\binom{N}{j}
			t_1^{N-j}(1-t_1)^j.	\label{eq:relevant_beta_binomial_relations}
		\end{aligned}
	\end{equation}
	The summands in the first and second relations are the binomial
	probability mass functions shown in panels (a) and (b) of
	Fig.~\ref{probability}, respectively. For the disjoint-region case
	($t_1<t_*<t_0$), the upper summation limit $Nt_*$ lies to the left
	of the mean $Nt_0$ of
	$X\sim{\rm Binomial}(N,t_0)$, while the upper summation limit
	$N(1-t_*)-1$ lies to the left of the mean $N(1-t_1)$ of
	$X\sim{\rm Binomial}(N,1-t_1)$. Thus, neither summation range crosses
	the mean of its corresponding binomial distribution, and both
	incomplete beta functions represent exponentially small lower-tail
	probabilities. When $N\xi\gg1$, their integral representations are
	dominated by the boundary points $t_0$ and $t_1$, respectively, and
	integration by parts gives
	\begin{equation}
		\begin{split}
			I_{1-t_0}\left((1-t_*)N, t_*N+1\right)\simeq\frac{\mathrm{e}^{-(N-1)D(y\|t_0)}\sqrt{(N-1)F}}{\sqrt{2\pi}(N-1)\partial_{t_0}D(y\|t_0)}-\frac{\partial_{t_0}^2D(y\|t_0)\mathrm{e}^{-(N-1)D(y\|t_0)}\sqrt{(N-1)F}}{\sqrt{2\pi}(N-1)^2\left(\partial_{t_0}D(y\|t_0)\right)^3 }\label{T0},
		\end{split}
	\end{equation}
	\begin{equation}
		\begin{split}
			I_{t_1}\left(t_*N+1,(1-t_*)N\right)\simeq-\frac{\mathrm{e}^{-(N-1)D(y\|t_1)}\sqrt{(N-1)F}}{\sqrt{2\pi}(N-1)\partial_{t_1}D(y\|t_1)}+\frac{\partial_{t_1}^2D(y\|t_1)\mathrm{e}^{-(N-1)D(y\|t_1)}\sqrt{(N-1)F}}{\sqrt{2\pi}(N-1)^2\left(\partial_{t_1}D(y\|t_1)\right)^3 }.\label{T1}
		\end{split}
	\end{equation}
	Substituting Eqs.~\eqref{T0} and \eqref{T1} into Eq.~\eqref{15} gives the minimum error probability
	\begin{equation}
		\begin{split}
			P_{\rm e}^{\rm min}
			&\simeq\sum_{i=0}^{1}\frac{2\sqrt{2\pi (1-t_*)}q(2t_i-1)t_i^2(1-t_i)}{\sqrt{t_*}(t_i-t_*)^2}N^{-\frac{3}{2}}\mathrm{e}^{-ND(t_*\|t_i)},
		\end{split}
	\end{equation}
	where $\xi=D(t_*\|t_0)=D(t_*\|t_1)$ is the Chernoff exponent~\cite{chernoff1952measure,mosonyi2021error}, and the coefficient $A_s$ introduced in the main text is
	explicitly given by
	$
	A_s
	=
	\sum_{i=0}^{1}
	[2\sqrt{2\pi(1-t_*)}\
	q(2t_i-1)t_i^2(1-t_i)]/
	[\sqrt{t_*}(t_i-t_*)^2].
	$
	
	We next consider the adjacent-region case $(t_0=t_1=t_*)$.  In this case, the cutoff coincides with the mean of the corresponding binomial distribution, and the two complementary incomplete beta functions approach $1/2$ in the large-$N$ limit, with finite-$N$ corrections of order $O(N^{-1/2})$. Using
	\begin{equation}
		\mathcal Q(t_i)-\mathcal Q(t_*)
		=
		q(2t_*-1)(t_i-t_*)+
		O\left((t_i-t_*)^2\right)
	\end{equation}
	and
	\begin{equation}
		\partial_tD(t_*\|t)|_{t=t_i}
		=
		\frac{t_i-t_*}{t_*(1-t_*)}
		+
		O\left((t_i-t_*)^2\right),
	\end{equation}
	we find that the two boundaries give identical contributions. Consequently,
	\begin{equation}
		\begin{split}
			P_{\rm e}^{\rm min}
			\simeq\frac{8\pi q(2t_*-1)}{\sqrt{2\pi N F}}.
		\end{split}
	\end{equation}
	Here the coefficient $A_b$ introduced in the main text is
	explicitly given by
	$A_b=4\sqrt{2\pi}\,q(2t_*-1)$.
	
	For the overlapping-region case ($t_0<t_*<t_1$), as illustrated in panels (c) and (d) of Fig.~\ref{probability}, the binomial cumulative distributions that correspond through the beta--binomial relation to
	$
	I_{1-t_0}((1-t_*)N,t_*N+1)
	$
	and
	$
	I_{t_1}(t_*N+1,(1-t_*)N)
	$
	extend beyond the means of their respective binomial distributions. Consequently, both cumulative probabilities approach unity,whereas the complementary probabilities
	$
	1-I_{1-t_0}\!\left((1-t_*)N,t_*N+1\right)
	$ and $
	1-I_{t_1}\!\left(t_*N+1,(1-t_*)N\right)
	$
	are exponentially small tail probabilities. 
	Expanding the two exponentially small complementary probabilities by the same boundary method as in the disjoint-region case gives
	\begin{equation}
		P_{\rm e}^{\rm min}
		\simeq
		P_{\rm ov}
		+
		A_{\rm s}N^{-3/2}\mathrm e^{-N\xi},
	\end{equation}
	where
	$
	\xi=D(t_*\|t_0)=D(t_*\|t_1)
	$
	and the intrinsic error probability introduced in the main text is
	$
	P_{\rm ov}
	=
	4\pi[\mathcal Q(t_1)-\mathcal Q(t_0)].
	$
	The intrinsic error arises from the overlap
	$t\in[t_0,t_1]$ between the admissible and rejected regions.
	
	\section{The Optimal Measurement and Minimum Error Probability for PQ}
	
	A single-qubit state with Bloch-vector length $r$ and direction $\boldsymbol n(\theta,\phi)$ can be written as
	\begin{equation}
		\rho_{r,\theta,\phi}
		=
		U_{\theta,\phi}D_rU_{\theta,\phi}^{\dagger},
	\end{equation}
	where
	$
	D_r=
	\begin{pmatrix}
		t & 0\\
		0 & 1-t
	\end{pmatrix},
	$
	with $t=(1+r)/2$, represents the state whose Bloch vector has
	length $r$ and points along the positive $z$ direction.
	We define the angularly averaged $N$-copy state as
	\begin{equation}
		\overline{\rho}^{(N)}_r
		\equiv
		\int_{S^2}{\rm d}\Omega\,
		\rho_{r,\theta,\phi}^{\otimes N}=4\pi\int_{SU(2)}{\rm d}\mu_{\mathrm H}(U)\,
		(U_{\theta,\phi}D_rU_{\theta,\phi}^{\dagger})^{\otimes N},
		\label{eq:PQ_angular_average}
	\end{equation}
	where ${\rm d}\mu_{\mathrm H}(U)$ denotes the Haar measure on $SU(2)$.
	
	The Schur--Weyl decomposition of the $N$-qubit Hilbert space
	decomposes is
	\begin{equation}
		\mathcal H_{1/2}^{\otimes N}
		=
		\bigoplus_{S=S_{\min}}^{S_{\max}}
		\mathcal V_S\otimes\mathcal K_S,
		\label{eq:PQ_Schur_Weyl}
	\end{equation}
	where $S_{\max}=N/2$, $S_{\min}=0$ for even $N$ and
	$S_{\min}=1/2$ for odd $N$. For each $S$, $\mathcal V_S$ carries
	the $(2S+1)$-dimensional irreducible representation of $SU(2)$,
	and $\mathcal K_S$ is the corresponding multiplicity space
	carrying an irreducible representation of the symmetric group
	$\mathcal S_N$~\cite{bartlett2007reference}.
	
	We choose a basis adapted to this decomposition,
	$
	\ket{S,M,\alpha}
	=
	\ket{S,M}_{\mathcal V_S}
	\otimes
	\ket{\alpha}_{\mathcal K_S},
	$
	where $M=-S,-S+1,\ldots,S$ labels the basis vectors of $\mathcal V_S$,
	whereas $\alpha=1,\ldots,d_S$ labels the basis vectors of $\mathcal K_S$.
	Here, $d_S=\binom{N}{N/2-S}-\binom{N}{N/2-S-1}$ is the multiplicity of the
	spin-$S$ representation~\cite{Pauncz1979}.
	The angular integration in Eq.~\eqref{eq:PQ_angular_average} makes
	$\overline{\rho}^{(N)}_r$ invariant under collective $SU(2)$
	rotations. In addition, since each
	$\rho_{r,\theta,\phi}^{\otimes N}$ is invariant under permutations
	of the $N$ copies, $\overline{\rho}^{(N)}_r$ is also invariant under
	the action of $\mathcal S_N$. Schur's lemma therefore implies that,
	within each spin-$S$ sector, $\overline{\rho}^{(N)}_r$ is
	proportional to the identity on both $\mathcal V_S$ and
	$\mathcal K_S$. Hence,
	\begin{equation}
		\begin{aligned}
			\overline{\rho}^{(N)}_r
			&=
			\bigoplus_{S=S_{\min}}^{S_{\max}}
			C_S(r)
			\left(
			\mathbbm 1_{\mathcal V_S}
			\otimes
			\mathbbm 1_{\mathcal K_S}
			\right)
			=
			\sum_{S=S_{\min}}^{S_{\max}}
			C_S(r)
			\sum_{M=-S}^{S}
			\sum_{\alpha=1}^{d_S}
			\ket{S,M,\alpha}\!\bra{S,M,\alpha}.
		\end{aligned}
		\label{eq:PQ_average_Schur_basis}
	\end{equation}

	To determine $C_S(r)$, we first note that
	\begin{equation}
		D_r^{\otimes N}	\ket{S,M,\alpha}
		=
		t^{\frac{N}{2}+M}
		(1-t)^{\frac{N}{2}-M}
		\ket{S,M,\alpha}.
	\end{equation}
	Let $\operatorname{Tr}_S$ denote the trace over the complete
	spin-$S$ sector $\mathcal V_S\otimes\mathcal K_S$. Taking this
	trace in Eq.~\eqref{eq:PQ_average_Schur_basis} gives
	$(2S+1)d_SC_S(r)$. On the other hand, taking the same trace in
	Eq.~\eqref{eq:PQ_angular_average} and using the invariance of the
	spin-$S$ projector under collective rotations gives
	$4\pi\operatorname{Tr}_S(D_r^{\otimes N})$. Equating the two
	results yields
	\begin{equation}
		\begin{split}
			(2S+1)d_SC_S(r)
			&=
			4\pi\operatorname{Tr}_S
			\left(D_r^{\otimes N}\right)
			=4\pi
			\left[
			\binom{N}{\frac{N}{2}-S}
			-
			\binom{N}{\frac{N}{2}-S-1}
			\right]
			\frac{t^{\frac{N}{2}-S}(1-t)^{\frac{N}{2}+S+1}-t^{\frac{N}{2}+S+1}(1-t)^{\frac{N}{2}-S}}{1-2t}.
			\label{31}
		\end{split}
	\end{equation}
	
	In this part, we obtain the optimal measurement for PQ by substituting a projective measurement characterized by a cutoff radius $x$ into the error probability and then imposing the stationarity condition. For the disjoint-region case ($r_0>r_1$), the projective measurement is defined as
	\begin{equation}
		\begin{split}
			E_0^{\rm PQ}(x)
			&=
			\bigoplus_{S=xS_{\rm max}+1}^{S_{\rm max}}  \left( \mathbbm{1}_{\mathcal{V}_S}\otimes \mathbbm{1}_{\mathcal{K}_S}\right).\label{E0pu}
		\end{split}
	\end{equation}
	Substituting Eq.~\eqref{E0pu} and $S_{\rm max}=N/2$ into the expression for the error probability
	\begin{equation}
		\begin{split}
			P_{\rm e}(x)
			=&
			4\pi\Bigg\{
			\sum_{S=S_{\min}}^{x\frac{N}{2}}
			\left[\binom{N}{\frac{N}{2}-S}
			-
			\binom{N}{\frac{N}{2}-S-1} \right] 
			\int_{r_0}^{1}
			w\left(r\right)\frac{t^{\frac{N}{2}-S}(1-t)^{\frac{N}{2}+S+1}-t^{\frac{N}{2}+S+1}(1-t)^{\frac{N}{2}-S}}{1-2t}{\rm d}r \\
			&+
			\sum_{S=x\frac{N}{2}+1 }^{N/2}
			\left[\binom{N}{\frac{N}{2}-S}
			-
			\binom{N}{\frac{N}{2}-S-1} \right] 
			\int_{0}^{r_1}
			w\left(r\right)\frac{t^{\frac{N}{2}-S}(1-t)^{\frac{N}{2}+S+1}-t^{\frac{N}{2}+S+1}(1-t)^{\frac{N}{2}-S}}{1-2t}{\rm d}r
			\Bigg\},	
		\end{split}
	\end{equation}

	For large $N$, the cutoff radius $x$ may be treated as a continuous variable. The stationarity condition
	$\left.{\rm d}P_{\rm e}(x)/{\rm d}x\right|_{x=x_*}=0$
	is, at leading order, equivalent to the equality of the two boundary contributions at the threshold $x=x_*$:
	\begin{equation}
		\int_{r_0}^{1}
		\frac{w(r)t^{\frac{N}{2}-S}(1-t)^{\frac{N}{2}+S+1}-t^{\frac{N}{2}+S+1}(1-t)^{\frac{N}{2}-S}}{1-2t} {\rm d}r
		=
		\int_{0}^{r_1}
		\frac{w(r)t^{\frac{N}{2}-S}(1-t)^{\frac{N}{2}+S+1}-t^{\frac{N}{2}+S+1}(1-t)^{\frac{N}{2}-S}}{1-2t} {\rm d}r .
	\end{equation}
	In the large-$N$ limit, the two integrals are dominated by the boundary points $r_0$ and $r_1$, respectively. Therefore the stationarity condition reduces, at the exponential order, to
	\begin{equation}
		t_0^{t_*N}(1-t_0)^{(1-t_*)N}
		=
		t_1^{t_*N}(1-t_1)^{(1-t_*)N},
	\end{equation}
	where $t_i=(1+r_i)/2$ and $t_*=(1+x_*)/2$. Taking the logarithm and solving for $x_*$ yields
	\begin{equation}
		x_*=\frac{\ln(1-r_1^2)-\ln({1-r_0^2})}{\ln\frac{1+r_0}{1-r_0}-\ln\frac{1+r_1}{1-r_1}}.\label{PQ_x_*}
	\end{equation}
	For the adjacent-region case
	($r_0=r_1$), the corresponding cutoff is obtained from the
	disjoint-region result by taking the limit $r_1\to r_0$.
	Applying this limit to Eq.~\eqref{PQ_x_*} yields
	\begin{equation}
		x_*
		=
		\lim_{r_1\rightarrow r_0}\frac{\ln(1-r_1^2)-\ln({1-r_0^2})}{\ln\frac{1+r_0}{1-r_0}-\ln\frac{1+r_1}{1-r_1}}=r_0.
	\end{equation}
	For the overlapping-region case
	($\theta_0>\theta_1$), the cutoff $x_*$ is given by the same
	expression as in the disjoint-region case.
	
	We next derive the asymptotic form of the minimum error probability for PQ.  Substituting the cutoff radius $x_*$ into the expression for the error probability gives the minimum error probability. The minimum error probability reduces to
	\begin{equation}
		\begin{split}
			P_{\rm e}^{\rm min}=&4\pi\bigg[\int_0^{r_1}{\rm d}rw(r)\frac{r(r-1)}{2}[I_{\frac{1-r}{2}}\left(t_*N+1,(1-t_*)N\right)+I_{\frac{1+r}{2}}\left(t_*N+2,(1-t_*)N-1\right) ]\\
			&+\int_0^{r_1}{\rm d}rw(r)\frac{r(r+1)}{2}[I_{\frac{1-r}{2}}\left(t_*N+2,(1-t_*)N-1\right)+I_{\frac{1+r}{2}}\left(t_*N+1,(1-t_*)N\right)]\\
			&+\int_{r_0}^1{\rm d}rw(r)\frac{r(r-1)}{2}[I_{\frac{1-r}{2}}\left((1-t_*)N-1,t_*N+2\right)-I_{\frac{1-r}{2}}\left(t_*N+1,(1-t_*)N\right)]\\
			&+\int_{r_0}^1{\rm d}rw(r)\frac{r(r+1)}{2}[I_{\frac{1-r}{2}}\left((1-t_*)N,t_*N+1\right)-I_{\frac{1-r}{2}}\left(t_*N+2,(1-t_*)N-1\right)]\bigg].
		\end{split}
	\end{equation}
	Among the eight regularized incomplete beta-function terms in the above
	expression, four are exponentially subleading compared with the other four. Retaining the remaining four terms,
	we obtain
	\begin{equation}
		\begin{split}
			P_{\rm e}^{\rm min}\simeq&4\pi\bigg[\int_0^{r_1}{\rm d}rw(r)\frac{r(r-1)}{2}I_{\frac{1+r}{2}}\left(t_*N+2,(1-t_*)N-1\right) +\int_0^{r_1}{\rm d}rw(r)\frac{r(r+1)}{2}I_{\frac{1+r}{2}}\left(t_*N+1,(1-t_*)N\right)\\
			&+\int_{r_0}^1{\rm d}rw(r)\frac{r(r-1)}{2}I_{\frac{1-r}{2}}\left((1-t_*)N-1,t_*N+2\right)+\int_{r_0}^1{\rm d}rw(r)\frac{r(r+1)}{2}I_{\frac{1-r}{2}}\left((1-t_*)N,t_*N+1\right)\bigg].
		\end{split}
	\end{equation}
	
	The subsequent large-$N$ asymptotic analysis follows the same procedure as that in Eqs.~\eqref{13}--\eqref{15} for PDQ, with only the prefactors modified. For the disjoint-region case,
	\begin{equation}
		\begin{split}
			P_{\rm e}^{\rm min}=&4\pi\sum_{i=0}^1\frac{w(r_i)r_it_i^2}{\sqrt{2\pi}(t_i-t_*)^2}\sqrt{\frac{1-t_*}{t_*}}\left[\frac{1-t_*}{t_*}t_i(r_i-1)+(1-t_i)(r_i+1)\right] N^{-\frac{3}{2}}\mathrm{e}^{-ND(t_*\|t_i)},
		\end{split}
	\end{equation}
	where
	$\xi=D(t_*\|t_0)=D(t_*\|t_1)$ is the Chernoff exponent, and the
	coefficient $A_s$ introduced in the main text is explicitly given by $A_s=4\pi\sum_{i=0}^1w(r_i)r_it_i^2\sqrt{1-t_*}[(1-t_*)t_i(r_i-1)/t_*+(1-t_i)(r_i+1)]/[ \sqrt{2\pi t_*}(t_i-t_*)^2] $. For the adjacent-region case, 
	\begin{equation}
		\begin{split}
			P_{\rm e}^{\rm min}\simeq&\frac{16\pi x_*^2w(x_*)}{\sqrt{2\pi N F}}.
		\end{split}
	\end{equation}
	Here
	$F=1/[t_*(1-t_*)]$ is the Fisher information of the Bernoulli
	distribution, and the coefficient $A_b$ introduced in the main text
	is explicitly given by $A_b=8\sqrt{2\pi}x_*^2w(x_*)$. For the overlapping-region case,
	\begin{equation}
		\begin{split}
			P_{\rm e}^{\rm min}\simeq&P_{\rm ov}+A_{\rm s} N^{-\frac{3}{2}}\mathrm{e}^{-ND(t_*\|t_i)}.
		\end{split}
	\end{equation}
	Here $P_{\rm ov}=4\pi\int_{r_0}^{r_1}w(r)r^2\,{\rm d}r$
	is the intrinsic error probability introduced in the main text.

	\section{Location of the Worst  Pairwise States}

	\subsection{A. The Worst Pairwise States in PDQ}
	In this subsection, we determine the locations of the worst pairwise
	states in PDQ. The search is performed within the convex hulls
	$\mathcal C_i$ of the two quantum-state families associated with the
	PDQ parameter regions $\Theta_i^{\rm PDQ}$ introduced in the main text.
	Specifically,
	\begin{equation}
		\begin{split}
			\mathcal C_i
			&=
			\operatorname{conv}
			\left\{
			\rho_{\boldsymbol\theta}:
			\boldsymbol\theta\in\Theta_i^{\rm PDQ}
			\right\}=
			\left\{
			\sum_{k=1}^{m}p_k\rho_{\boldsymbol\theta_k}:
			\boldsymbol\theta_k\in\Theta_i^{\rm PDQ},\
			p_k\geq0,\
			\sum_{k=1}^{m}p_k=1,\
			m\in\mathbb N
			\right\},
			\qquad i=0,1.
		\end{split}
	\end{equation}
	Here, the PDQ parameter regions are defined by
	\begin{equation}
		\begin{split}
			\Theta_{0}^{\rm PDQ}
			&=
			\{(\theta,\phi):0\leq\theta\leq\theta_0,\;0\leq\phi<2\pi\},\qquad
			\Theta_{1}^{\rm PDQ}
			=
			\{(\theta,\phi):\theta_1\leq\theta\leq\pi,\;0\leq\phi<2\pi\}.\label{polarization}
		\end{split}
	\end{equation}
	Within these convex hulls, the worst pairwise states are defined as a pair of states, one from each convex hull, that minimizes the quantum Chernoff divergence over all possible pairs. Here, the quantum Chernoff divergence is defined as~\cite{audenaert2007discriminating}
	\begin{equation}
		C_{\rm Q}\left(\rho_0\|\rho_1\right)
		=
		-\ln\left[
		\min_{0\leq s\leq1}
		\operatorname{Tr}
		\left(\rho_0^s\rho_1^{1-s}\right)
		\right]=-\ln\left[
		\min_{0\leq s\leq1}
		T_s(\rho_0,\rho_1)
		\right],
	\end{equation}
	where we define $T_s(\rho_0,\rho_1)=\operatorname{Tr}
	\left(\rho_0^s\rho_1^{1-s}\right)$.
	Thus, the worst pairwise states $\rho_{{\rm w},0}$ and $\rho_{{\rm w},1}$ satisfy~\cite{mosonyi2021error}
	\begin{equation}
		C_{\rm Q}\left(
		\rho_{{\rm w},0}\|
		\rho_{{\rm w},1}
		\right)
		=\min_{\substack{
				\rho_0\in\mathcal C_0,
				\rho_1\in\mathcal C_1}}\left\{-\ln\left[
		\min_{0\leq s\leq1}
		T_s(\rho_0,\rho_1)
		\right]\right\}=
		-\ln\left[
		\max_{\substack{
				\rho_0\in\mathcal C_0,
				\rho_1\in\mathcal C_1}}
		\min_{0\leq s\leq1}
		T_s(\rho_0,\rho_1)
		\right].
	\end{equation}
	Equivalently,
	\begin{equation}
		\left(
		\rho_{{\rm w},0},
		\rho_{{\rm w},1}
		\right)
		=
		\underset{
			\rho_0\in\mathcal C_0,\,
			\rho_1\in\mathcal C_1
		}{
			\operatorname{arg\,min}
		}
		C_{\rm Q}(\rho_0\|\rho_1)=\underset{
			\rho_0\in\mathcal C_0,\,
			\rho_1\in\mathcal C_1
		}{
			\operatorname{arg\,max}
		}\min_{0\leq s\leq1}
		T_s(\rho_0,\rho_1)
		.
	\end{equation}

	\noindent\textbf{Proposition 1 (Minimax equality in PDQ).---}
	The maximin and minimax values of
	$T_s(\rho_0,\rho_1)$ coincide:
	\begin{equation}
		\max_{\substack{
				\rho_0\in\mathcal C_0,
				\rho_1\in\mathcal C_1}}
		\min_{0\leq s\leq1}
		T_s(\rho_0,\rho_1)
		=
		\min_{0\leq s\leq1}
		\max_{\substack{
				\rho_0\in\mathcal C_0,
				\rho_1\in\mathcal C_1}}
		T_s(\rho_0,\rho_1) .
		\label{eq:PDQ_minimax}
	\end{equation}
	\noindent\textit{Proof.---}
	Let
	$
	\mathcal X
	=
	\mathcal C_0\times\mathcal C_1
	$ and $
	\mathcal Y=[0,1]
	$. Since $\mathcal C_0$ and $\mathcal C_1$ are compact convex sets,
	$\mathcal X$ is also compact and convex, and $\mathcal Y=[0,1]$
	is a compact convex interval.
	
	For every fixed $s\in[0,1]$, Lieb's concavity theorem~\cite{LIEB1973267} shows that
	$T_s(\rho_0,\rho_1)$ is jointly concave in
	$(\rho_0,\rho_1)$. It is therefore concave and upper
	semicontinuous on $\mathcal X$.
	
	For every fixed pair $(\rho_0,\rho_1)$, consider the spectral
	decompositions
	\begin{equation}
		\rho_0=\sum_a\lambda_aP_{0,a},
		\qquad
		\rho_1=\sum_b\mu_bP_{1,b }.
	\end{equation}
	Then
	\begin{equation}
		T_s(\rho_0,\rho_1)
		=
		\sum_{a,b}
		\lambda_a^s\mu_b^{1-s}
		\operatorname{Tr}(P_{0,a}P_{1,b}).
	\end{equation}
	Each summand in this sum is a nonnegative multiple of an exponential function $s$
	and is therefore convex in $s$. Hence,
	$T_s(\rho_0,\rho_1)$ is convex and lower semicontinuous on
	$\mathcal Y$.

	All the conditions of Sion's minimax theorem~\cite{Sion1958} are
	therefore satisfied. It follows that
	\begin{equation}
		\max_{\substack{
				\rho_0\in\mathcal C_0,
				\rho_1\in\mathcal C_1}}
		\min_{0\leq s\leq1}
		T_s(\rho_0,\rho_1)
		=
		\min_{0\leq s\leq1}
		\max_{\substack{
				\rho_0\in\mathcal C_0,
				\rho_1\in\mathcal C_1}}
		T_s(\rho_0,\rho_1).
	\end{equation}

	\noindent\textbf{Proposition 2 (worst pairwise states in PDQ).---}
	For every fixed $s\in(0,1)$, the maximum of
	$T_s(\rho_0,\rho_1)$ over
	$(\rho_0,\rho_1)\in\mathcal C_0\times\mathcal C_1$
	is attained by the worst pairwise states, whose representations in
	the $S_z$ eigenbasis are
	\begin{equation}
		\rho_{{\rm w},0}
		=
		\begin{pmatrix}
			\dfrac{1+\cos\theta_0}{2} & 0\\
			0 & \dfrac{1-\cos\theta_0}{2}
		\end{pmatrix},
		\qquad
		\rho_{{\rm w},1}
		=
		\begin{pmatrix}
			\dfrac{1+\cos\theta_1}{2} & 0\\
			0 & \dfrac{1-\cos\theta_1}{2}
		\end{pmatrix}.
		\label{eq:PDQ_worst_pair}
	\end{equation}

	\noindent\textit{Proof.---}
	Owing to the geometric properties of the PDQ convex hulls, the Bloch vector of any state $\rho_i\in\mathcal C_i$ can be projected onto the $z$ axis, with the resulting state remaining in the same convex hull. Accordingly, for an arbitrary pair $\rho_0\in\mathcal C_0$ and $\rho_1\in\mathcal C_1$, we define the projected states
	\begin{equation}
		\widetilde{\rho}_i
		=
		\frac{1}{2}
		\left(
		\rho_i+\sigma_z\rho_i\sigma_z
		\right)
		=
		\frac{\mathbbm 1+z_i\sigma_z}{2},
		\qquad i=0,1,
	\end{equation}
	where $z_i$ is the $z$ component of the Bloch vector of $\rho_i$.

	By Lieb's joint concavity theorem~\cite{LIEB1973267},
	$T_s(\rho_0,\rho_1)$ is jointly concave in
	$(\rho_0,\rho_1)$ for every fixed $s\in(0,1)$. Accordingly,
	\begin{equation}
		T_s\left(
		\widetilde{\rho}_0,\widetilde{\rho}_1
		\right)
		=
		T_s
		\left(
		\frac{1}{2}\rho_0+\frac{1}{2}\sigma_z\rho_0\sigma_z,
		\frac{1}{2}\rho_1+\frac{1}{2}\sigma_z\rho_1\sigma_z
		\right)
		\geq
		\frac{1}{2}T_s(\rho_0,\rho_1)
		+
		\frac{1}{2}
		T_s
		(\sigma_z\rho_0\sigma_z,
		\sigma_z\rho_1\sigma_z)
		=
		T_s(\rho_0,\rho_1),
	\end{equation}
	where the last equality follows from the invariance of $T_s(\rho_0,\rho_1)$ under the simultaneous transformation of both states by $\sigma_z$. Explicitly, using $\sigma_z^\dagger=\sigma_z$ and $\sigma_z^2=\mathbbm 1$, we have 
	\begin{equation}
		T_s(\sigma_z\rho_0\sigma_z,\sigma_z\rho_1\sigma_z)={\rm Tr}\left[(\sigma_z\rho_0\sigma_z)^s(\sigma_z\rho_1\sigma_z)^{1-s}\right]={\rm Tr}\left(\sigma_z\rho_0^s\rho_1^{1-s}\sigma_z\right)={\rm Tr}\left(\rho_0^s\rho_1^{1-s}\right)=T_s(\rho_0,\rho_1).
	\end{equation}

	Therefore, in maximizing $T_s(\rho_0,\rho_1)$, it is
	sufficient to consider pairs of states whose Bloch vectors lie on the
	$z$ axis.
	For states whose Bloch vectors lie on the $z$ axis,
	\begin{equation}
		T_s(\rho_0,\rho_1)
		=
		\left(\frac{1+z_0}{2}\right)^s
		\left(\frac{1+z_1}{2}\right)^{1-s}+
		\left(\frac{1-z_0}{2}\right)^s
		\left(\frac{1-z_1}{2}\right)^{1-s},
	\end{equation}
	where
	$z_0\in[\cos\theta_0,1]$ and
	$z_1\in[-1,\cos\theta_1]$. Taking the partial derivatives with
	respect to $z_0$ and $z_1$ gives
	\begin{equation}
		\frac{
			\partial T_s(\rho_0,\rho_1)
		}{
			\partial z_0
		}
		=
		\frac{s}{2}
		\left[
		\left(
		\frac{1+z_1}{1+z_0}
		\right)^{1-s}
		-
		\left(
		\frac{1-z_1}{1-z_0}
		\right)^{1-s}
		\right]
		<0,
	\end{equation}
	and
	\begin{equation}
		\frac{
			\partial T_s(\rho_0,\rho_1)
		}{
			\partial z_1
		}
		=
		\frac{1-s}{2}
		\left[
		\left(
		\frac{1+z_0}{1+z_1}
		\right)^s
		-
		\left(
		\frac{1-z_0}{1-z_1}
		\right)^s
		\right]
		>0.
	\end{equation}
	Thus, $T_s(\rho_0,\rho_1)$ decreases with $z_0$ and
	increases with $z_1$. Its maximum is attained at
	$
	z_0=\cos\theta_0,
	$ $
	z_1=\cos\theta_1.
	$
	The worst pairwise states are consequently
	\begin{equation}
		\rho_{{\rm w},0}
		=
		\begin{pmatrix}
			\dfrac{1+\cos\theta_0}{2} & 0\\
			0 & \dfrac{1-\cos\theta_0}{2}
		\end{pmatrix},
		\qquad
		\rho_{{\rm w},1}
		=
		\begin{pmatrix}
			\dfrac{1+\cos\theta_1}{2} & 0\\
			0 & \dfrac{1-\cos\theta_1}{2}
		\end{pmatrix}.
	\end{equation}
	
	\noindent\textbf{Inference 3 (the Chernoff divergence of the worst pairwise states in PDQ).---}
	The quantum Chernoff divergence of the worst pairwise states in PDQ
	is equal to the Chernoff exponent governing the minimum error
	probability derived in the main text; namely,
	\begin{equation}
		C_{\rm Q}\left(
		\rho_{{\rm w},0}\|
		\rho_{{\rm w},1}
		\right)
		=
		-\ln\left[
		\min_{0\leq s\leq1}
		T_s(\rho_{{\rm w},0},\rho_{{\rm w},1})
		\right]
		=
		\xi,
		\label{eq:PDQ_QCB_equals_exponent}
	\end{equation}
	where $\xi$ denotes the Chernoff exponent of
	$P_\mathrm{e}^{\min}$ obtained in the main text.
	
	\noindent\textit{Proof.---}
	For the worst pairwise states given in
	Eq.~\eqref{eq:PDQ_worst_pair}, let
	$
	t_i=(1+\cos\theta_i)/2.
	$
	The function
	\begin{equation}
		T_s(\rho_{{\rm w},0},\rho_{{\rm w},1})
		=
		t_0^s t_1^{1-s}
		+
		(1-t_0)^s(1-t_1)^{1-s}.
		\label{eq:PDQ_worst_pair_T}
	\end{equation}
	
	To determine the minimum of
	$T_s(\rho_{{\rm w},0},\rho_{{\rm w},1})$ over
	$0<s<1$, we differentiate it with respect to $s$. The stationarity
	condition at $s=s_*$ is
	\begin{equation}
		\left.
		\frac{
			\partial
			T_s(\rho_{{\rm w},0},\rho_{{\rm w},1})
		}{
			\partial s
		}
		\right|_{s=s_*}
		=
		t_0^{s_*}t_1^{1-s_*}
		\ln\frac{t_0}{t_1}
		+
		(1-t_0)^{s_*}(1-t_1)^{1-s_*}
		\ln\frac{1-t_0}{1-t_1}
		=0.
		\label{eq:PDQ_QCB_stationarity}
	\end{equation}
	Moreover,
	\begin{equation}
		\left.\frac{
			\partial^2
			T_s(\rho_{{\rm w},0},\rho_{{\rm w},1})
		}{
			\partial s^2
		}\right|_{s=s_*}
		=
		t_0^s t_1^{1-s}
		\left(
		\ln\frac{t_0}{t_1}
		\right)^2
		+
		(1-t_0)^s(1-t_1)^{1-s}
		\left(
		\ln\frac{1-t_0}{1-t_1}
		\right)^2
		>0.
	\end{equation}
	Therefore,
	$T_s(\rho_0,\rho_1)$ is
	strictly convex in $s$, and the stationary point $s_*$ is its unique
	global minimizer.
	
	Combining Eq.~\eqref{sex_*} with the stationarity condition in
	Eq.~\eqref{eq:PDQ_QCB_stationarity}, we obtain
	\begin{equation}
		t_*
		=
		\frac{
			t_0^{s_*}t_1^{1-s_*}
		}{
			T_s(\rho_{{\rm w},0},\rho_{{\rm w},1})
		},\qquad	1-t_*
		=
		\frac{
			(1-t_0)^{s_*}(1-t_1)^{1-s_*}
		}{
			T_s(\rho_{{\rm w},0},\rho_{{\rm w},1})
		}.
		\label{eq:PDQ_tstar_from_sstar}
	\end{equation}
	It therefore follows that
	\begin{equation}
		C_{\rm Q}\left(
		\rho_{{\rm w},0}\|
		\rho_{{\rm w},1}
		\right)
		=
		\ln
		\frac{
			t_*
		}{
			t_0^{s_*}t_1^{1-s_*}
		}
		=
		\ln
		\frac{
			1-t_*
		}{
			(1-t_0)^{s_*}(1-t_1)^{1-s_*}
		}.
		\label{eq:PDQ_log_T_relation}
	\end{equation}
	Multiplying the first logarithmic expression in
	Eq.~\eqref{eq:PDQ_log_T_relation} by $t_*$ and the second by
	$1-t_*$, and then adding the two expressions, we obtain
	\begin{equation}
		C_{\rm Q}\left(
		\rho_{{\rm w},0}\|
		\rho_{{\rm w},1}
		\right)
		=
		s_*D(t_*\|t_0)
		+
		(1-s_*)D(t_*\|t_1).
		\label{eq:PDQ_QCB_KL_combination}
	\end{equation}
	Since Eq.~\eqref{sex_*} also gives
	$
	D(t_*\|t_0)=D(t_*\|t_1),
	$
	we finally obtain
	\begin{equation}
		C_{\rm Q}\left(
		\rho_{{\rm w},0}\|
		\rho_{{\rm w},1}
		\right)
		=
		D(t_*\|t_0)
		=
		D(t_*\|t_1)
		=
		\xi,
	\end{equation}
	where the last equality follows from the Chernoff exponent derived
	in the main text.

	\subsection{B. The Worst Pairwise States in PQ}
	
	In this subsection, we determine the locations of the worst pairwise
	states in PQ. The PQ parameter regions introduced in the main text
	are
	\begin{equation}
		\begin{split}
			\Theta_0^{\rm PQ}
			&=
			\left\{
			(r,\theta,\phi):
			r_0\leq r\leq1,\;
			0\leq\theta\leq\pi,\;
			0\leq\phi<2\pi
			\right\},\quad
			\Theta_1^{\rm PQ}
			=
			\left\{
			(r,\theta,\phi):
			0\leq r\leq r_1,\;
			0\leq\theta\leq\pi,\;
			0\leq\phi<2\pi
			\right\}.
		\end{split}
		\label{eq:PQ_parameter_regions}
	\end{equation}
	The search is performed directly within the corresponding original
	quantum-state families
	\begin{equation}
		\mathcal S_i
		=
		\left\{
		\rho_{r,\theta,\phi}:
		(r,\theta,\phi)\in\Theta_i^{\rm PQ}
		\right\},
		\qquad i=0,1.
	\end{equation}
	
	\noindent\textbf{Proposition 4 ($s$-independent maximizers of
		$T_s$ in PQ).---}
	For every fixed $s\in(0,1)$, the state pairs that maximize
	$T_s(\rho_0,\rho_1)$ over
	$(\rho_0,\rho_1)\in\mathcal S_0\times\mathcal S_1$ are not unique.
	Moreover, the complete set of state pairs maximizing $T_s$ is the
	same for all $s\in(0,1)$ and is given by
	\begin{equation}
		\operatorname*{arg\,max}_{\substack{
				\rho_0\in\mathcal S_0,\,
				\rho_1\in\mathcal S_1}}
		T_s(\rho_0,\rho_1)
		=
		\left\{
		\left(
		\rho_{{\rm w},0}(\boldsymbol n),
		\rho_{{\rm w},1}(\boldsymbol n)
		\right):
		\boldsymbol n\in S^2
		\right\},
		\qquad s\in(0,1).
		\label{eq:PQ_common_maximizer_set}
	\end{equation}
	Here,
	\begin{equation}
		\rho_{{\rm w},0}(\boldsymbol n)
		=
		\frac{
			\mathbbm 1+r_0\boldsymbol n\cdot\boldsymbol\sigma
		}{2},
		\qquad
		\rho_{{\rm w},1}(\boldsymbol n)
		=
		\frac{
			\mathbbm 1+r_1\boldsymbol n\cdot\boldsymbol\sigma
		}{2}.
		\label{eq:PQ_worst_pair}
	\end{equation}
	\noindent\textit{Proof.---}
	Let
	\begin{equation}
		\rho_0
		=
		\frac{\mathbbm 1+r\boldsymbol n\cdot\boldsymbol\sigma}{2},
		\qquad
		\rho_1
		=
		\frac{\mathbbm 1+r'\boldsymbol n'\cdot\boldsymbol\sigma}{2},
	\end{equation}
	where
	$
	r\in[r_0,1]
	$,
	$
	r'\in[0,r_1]
	$,
	and $r_0>r_1$. Let $\gamma\in[0,\pi]$ be the angle between
	$\boldsymbol n$ and $\boldsymbol n'$. Defining
	$
	\lambda_\pm=(1\pm r)/2
	$
	and
	$
	\mu_\pm=(1\pm r')/2,
	$
	we obtain
	\begin{equation}
		\begin{split}
			T_s(\rho_0,\rho_1)
			&=
			\frac{1+\cos\gamma}{2}
			\left[
			\left(\frac{1+r}{2}\right)^s\left(\frac{1+r'}{2}\right)^{1-s}
			+
			\left(\frac{1-r}{2}\right)^s\left(\frac{1-r'}{2}\right)^{1-s}
			\right]\\
			&+
			\frac{1-\cos\gamma}{2}
			\left[
			\left(\frac{1+r}{2}\right)^s\left(\frac{1-r'}{2}\right)^{1-s}
			+
			\left(\frac{1-r}{2}\right)^s\left(\frac{1+r'}{2}\right)^{1-s}
			\right].
			\label{eq:PQ_T_general}
		\end{split}
	\end{equation}
	We first optimize
	$T_s(\rho_0,\rho_1)$ over the relative direction of
	the two Bloch vectors. Differentiating it with respect to $\gamma$
	gives
	\begin{equation}
		\frac{\partial
			T_s(\rho_0,\rho_1)}{\partial\gamma}
		=
		\frac{\sin\gamma}{2}
		\left[\left(\frac{1+r}{2}\right)^s-\left(\frac{1-r}{2}\right)^s\right]
		\left[\left(\frac{1-r'}{2}\right)^{1-s}-\left(\frac{1+r'}{2}\right)^{1-s}\right]\leq 0.
		\label{eq:PQ_derivative_gamma}
	\end{equation}
	The stationarity condition
	$\partial T_s(\rho_0,\rho_1)/\partial\gamma=0$
	gives
	$
	\gamma=0,
	\gamma=\pi 
	$ or $ 
	r'=0.
	$
	For $r'>0$, the inequality in
	Eq.~\eqref{eq:PQ_derivative_gamma} is strict for every
	$\gamma\in(0,\pi)$. Hence, any state pair maximizing
	$T_s(\rho_0,\rho_1)$ must have parallel Bloch vectors,
	namely $\gamma=0$. When $r'=0$, the state
	$\rho_1=\mathbbm 1/2$ is independent of its Bloch-vector direction,
	so the maximizing pair can again be represented by choosing
	$\gamma=0$.
	
	Having optimized over the directions of the Bloch vectors, we next
	determine the locations of the state pairs maximizing $T_s$ by optimizing
	over their lengths. Substituting $\gamma=0$ in
	Eq.~\eqref{eq:PQ_T_general} gives
	\begin{equation}
		T_s(\rho_0,\rho_1)
		=
		\left(\frac{1+r}{2}\right)^s
		\left(\frac{1+r'}{2}\right)^{1-s}
		+
		\left(\frac{1-r}{2}\right)^s
		\left(\frac{1-r'}{2}\right)^{1-s}.
	\end{equation}
	For a fixed $s\in(0,1)$, the partial derivatives
	with respect to $r$ and $r'$ are
	\begin{equation}
		\frac{\partial
			T_s(\rho_0,\rho_1)}{\partial r}
		=
		\frac{s}{2}
		\left[
		\left(\frac{1+r'}{1+r}\right)^{1-s}
		-
		\left(\frac{1-r'}{1-r}\right)^{1-s}
		\right]
		<0,\label{par}
	\end{equation}
	and
	\begin{equation}
		\frac{\partial
			T_s(\rho_0,\rho_1)}{\partial r'}
		=
		\frac{1-s}{2}
		\left[
		\left(\frac{1+r}{1+r'}\right)^s
		-
		\left(\frac{1-r}{1-r'}\right)^s
		\right]
		>0.\label{par'}
	\end{equation}
	Thus, $T_s(\rho_0,\rho_1)$ decreases with $r$ and increases with
	$r'$. Its maximum over the allowed ranges is therefore attained at
	$
	r=r_0
	$
	and
	$
	r'=r_1.
	$
	Combining these conditions with $\gamma=0$, we conclude that every
	state pair maximizing $T_s(\rho_0,\rho_1)$ is of the form
	\begin{equation}
		\rho_{{\rm w},0}(\boldsymbol n)
		=
		\frac{
			\mathbbm 1+r_0\boldsymbol n\cdot\boldsymbol\sigma
		}{2},
		\qquad
		\rho_{{\rm w},1}(\boldsymbol n)
		=
		\frac{
			\mathbbm 1+r_1\boldsymbol n\cdot\boldsymbol\sigma
		}{2},
		\qquad
		\boldsymbol n\in S^2.
		\label{eq:PQ_maximizing_pairs}
	\end{equation}
	Although the derivatives with respect to $\gamma$, $r$, and $r'$
	depend explicitly on $s$, their signs remain unchanged for all
	$s\in(0,1)$. Hence, the maximizing conditions
	$
	\gamma=0,
	$
	$
	r=r_0,
	$
	and
	$
	r'=r_1
	$
	are independent of $s$. It follows that
	\begin{equation}
		\operatorname*{arg\,max}_{\substack{
				\rho_0\in\mathcal S_0, 
				\rho_1\in\mathcal S_1}}
		T_s(\rho_0,\rho_1)
		=
		\left\{
		\left(
		\rho_{{\rm w},0}(\boldsymbol n),
		\rho_{{\rm w},1}(\boldsymbol n)
		\right):
		\boldsymbol n\in S^2
		\right\},
		\qquad s\in(0,1).
		\label{eq:PQ_common_maximizing_set}
	\end{equation}

	\noindent\textbf{Proposition 5 (Minimax equality in PQ).---}
	The maximin and minimax values of
	$T_s(\rho_0,\rho_1)$ coincide:
	\begin{equation}
		\max_{\substack{
				\rho_0\in\mathcal S_0,
				\rho_1\in\mathcal S_1}}
		\min_{0\leq s\leq1}
		T_s(\rho_0,\rho_1)
		=
		\min_{0\leq s\leq1}
		\max_{\substack{
				\rho_0\in\mathcal S_0,
				\rho_1\in\mathcal S_1}}
		T_s(\rho_0,\rho_1) .
		\label{eq:PQ_minimax}
	\end{equation}

	\noindent\textit{Proof.---}
	Let
	$
	\mathcal X=\mathcal S_0\times\mathcal S_1
	$
	denote the set of state pairs
	$(\rho_0,\rho_1)$ with
	$\rho_0\in\mathcal S_0$ and $\rho_1\in\mathcal S_1$.
	By Proposition~4 and the direct evaluations at $s=0$ and $s=1$,
	every state pair
	$
	(\rho_{0,*},\rho_{1,*})
	\in
	\left\{
	\left(
	\rho_{{\rm w},0}(\boldsymbol n),
	\rho_{{\rm w},1}(\boldsymbol n)
	\right):
	\boldsymbol n\in S^2
	\right\}
	$
	maximizes $T_s$ over $\mathcal X$ for every $s\in[0,1]$.
	Therefore, for every
	$(\rho_0,\rho_1)\in\mathcal X$ and every $s\in[0,1]$,
	\begin{equation}
		T_s(\rho_0,\rho_1)
		\leq
		T_s(\rho_{0,*},\rho_{1,*})
		=
		\max_{(\rho_0,\rho_1)\in\mathcal X}
		T_s(\rho_0,\rho_1).
		\label{eq:PQ_pointwise_upper_bound}
	\end{equation}
	Since the same pair $(\rho_{0,*},\rho_{1,*})$ maximizes $T_s$
	for every $s$, taking the minimum over $s$ gives
	\begin{equation}
		\min_{0\leq s\leq1}
		T_s(\rho_0,\rho_1)
		\leq
		\min_{0\leq s\leq1}
		\max_{(\rho_0,\rho_1)\in\mathcal X}
		T_s(\rho_0,\rho_1).
	\end{equation}
	Taking the maximum over
	$(\rho_0,\rho_1)\in\mathcal X$ on both sides, and noting that the
	right-hand side is independent of $(\rho_0,\rho_1)$, we obtain
	\begin{equation}
		\max_{(\rho_0,\rho_1)\in\mathcal X}
		\min_{0\leq s\leq1}
		T_s(\rho_0,\rho_1)
		\leq
		\min_{0\leq s\leq1}
		\max_{(\rho_0,\rho_1)\in\mathcal X}
		T_s(\rho_0,\rho_1).
		\label{eq:PQ_minimax_upper_bound}
	\end{equation}
	
	Because $(\rho_{0,*},\rho_{1,*})\in\mathcal X$ is an $s$-independent maximizing pair of $T_s(\rho_0,\rho_1)$ over $\mathcal X$, we have 
	\begin{equation}
		\begin{split}
			\max_{(\rho_0,\rho_1)\in\mathcal X}
			\min_{0\leq s\leq1}
			T_s(\rho_0,\rho_1)
			&\geq
			\min_{0\leq s\leq1}
			T_s(\rho_{0,*},\rho_{1,*})
			=
			\min_{0\leq s\leq1}
			\max_{(\rho_0,\rho_1)\in\mathcal X}
			T_s(\rho_0,\rho_1).
		\end{split}
		\label{eq:PQ_maximin_lower_bound}
	\end{equation}
	Combining Eqs.~\eqref{eq:PQ_minimax_upper_bound} and
	\eqref{eq:PQ_maximin_lower_bound}, we obtain
	\begin{equation}
		\max_{(\rho_0,\rho_1)\in\mathcal X}
		\min_{0\leq s\leq1}
		T_s(\rho_0,\rho_1)
		=
		\min_{0\leq s\leq1}
		\max_{(\rho_0,\rho_1)\in\mathcal X}
		T_s(\rho_0,\rho_1).
		\label{eq:PQ_maximin_common_pair}
	\end{equation}

	\noindent\textbf{Inference 6 (Worst pairwise states in PQ).---}
	Every state pair
	$
	(\rho_{0,*},\rho_{1,*})\in\left\{\left(\rho_{{\rm w},0}(\boldsymbol n),\rho_{{\rm w},1}(\boldsymbol n)\right):\boldsymbol n\in S^2\right\}
	$
	is one of the worst pairwise states in PQ and minimizes the quantum Chernoff divergence over $\mathcal S_0\times\mathcal S_1$.
	
	\noindent\textit{Proof.---}This result follows directly from Propositions~4 and 5.\hfill$\square$

	\noindent\textbf{Inference 7 (The Chernoff divergence of the worst
		pairwise states in PQ).---}
	The quantum Chernoff divergence of the worst pairwise states in PQ
	is equal to the Chernoff exponent governing the minimum error
	probability derived in the main text; namely,
	\begin{equation}
		C_{\rm Q}\left(
		\rho_{{\rm w},0}\|
		\rho_{{\rm w},1}
		\right)
		=
		-\ln\left[
		\min_{0\leq s\leq1}
		T_s(\rho_{{\rm w},0},\rho_{{\rm w},1})
		\right]
		=
		\xi,
		\label{eq:PQ_QCB_equals_exponent}
	\end{equation}
	where $\xi$ denotes the Chernoff exponent of
	$P_\mathrm{e}^{\min}$ obtained in the main text.
	
	\noindent\textit{Proof.---}For the worst pairwise states in Eq.~\eqref{eq:PQ_worst_pair}, let
	$t_i=(1+r_i)/2$, with $i=0,1$. Since the two states have parallel
	Bloch vectors, they commute, and their Chernoff coefficient is
	\begin{equation}
		T_s(\rho_{{\rm w},0},\rho_{{\rm w},0})
		=
		t_0^s t_1^{1-s}
		+
		(1-t_0)^s(1-t_1)^{1-s}.
	\end{equation}
	This expression has the same form as its PDQ counterpart.
	Therefore, applying the same argument as in the proof of Inference~3 establishes the claimed equality.

	\section{Scaling Function in PDQ}
	
	Near the critical point, we parameterize $t_i=t_*+\delta_i$, where $\delta_0=\delta$ and $\delta_1=-\delta$. Thus, for the disjoint-region case ($\delta>0$) near the critical point, the minimum error probability takes the asymptotic form
	\begin{equation}
		\begin{split}
			P_{\rm e}^{\rm min}=&A_{\rm s}N^{-\frac{3}{2}}\mathrm{e}^{-N\xi}\\
			=&\sum_{i=0}^{1}\frac{2\sqrt{2\pi (1-t_*)}q(2t_i-1)t_i^2(1-t_i)}{\sqrt{t_*}(t_i-t_*)^2}N^{-\frac{3}{2}}\mathrm{e}^{-N\xi}\\
			\simeq&\frac{2\sqrt{2\pi (1-t_*)t_*}q(2t_*-1)}{\xi}N^{-\frac{3}{2}}\mathrm{e}^{-N\xi}\\
			=&\frac{2\sqrt{2\pi (1-t_*)t_*}q(2t_*-1)}{\sqrt{N}}\frac{\mathrm{e}^{-N\xi}}{N\xi}\\
			=&\frac{\mathcal{C}_s(t_*)}{\sqrt{N}}f_{\rm s}(N\xi).
		\end{split}
	\end{equation}
	Here, the third line follows from the expansions
	\begin{equation}
		\begin{split}
			&q(2t_i-1)\simeq q(2t_*-1),\\
			&t_i^2(1-t_i)\simeq t_*^2(1-t_*),\\
			&\xi=D(t_*\|t_i)\simeq\frac{\delta^2}{2t_*(1-t_*)},
		\end{split}
	\end{equation}
	together with $t_i-t_*=\delta_i$. In the last line, we have defined $\mathcal{C}_{\rm s}(t_*)=2q(2t_*-1)\sqrt{2\pi t_*(1-t_*)}$ and $f_{\rm s}(a)={\rm e}^{-a}/a$ for the disjoint-region case.

	For the overlapping-region case ($\delta<0$) near the critical point, the minimum error probability takes the asymptotic form
	\begin{equation}
		\begin{split}
			P_{\rm e}^{\rm min}
			\simeq&P_{\rm ov}+A_{\rm s}N^{-3/2}{\rm e}^{-N\xi}.
		\end{split}
	\end{equation}
	Since $\xi$ is quadratic in $\delta$ near the critical point, the second term retains the same scaling form as in the disjoint-region case,
	\begin{equation}
		A_{\rm s}N^{-3/2}\mathrm{e}^{-N\xi}
		\simeq
		\frac{\mathcal C_{\rm s}(t_*)}{\sqrt N}f_{\rm s}(N\xi).
	\end{equation}
	For the intrinsic error probability 
	$P_{\rm ov}=4\pi[\mathcal Q(t_1)-\mathcal Q(t_0)]$,
	an expansion around $t_*$ gives
	\begin{equation}
		\begin{split}
			P_{\rm ov}
			&\simeq
			-8\pi q(2t_*-1)\delta\\
			&\simeq
			8\pi q(2t_*-1)\sqrt{2t_*(1-t_*)\xi}\\
			&=\frac{8\pi q(2t_*-1)\sqrt{2t_*(1-t_*)}}{\sqrt{N}}\sqrt{N\xi } \\
			&=
			\frac{\mathcal C_{\rm ov}(t_*)}{\sqrt N}f_{\rm ov}(N\xi),
		\end{split}
	\end{equation}
	where
	$
	\mathcal C_{\rm ov}(t_*)
	=
	8\pi q(2t_*-1)\sqrt{2t_*(1-t_*)}
	$ and $
	f_{\rm ov}(a)=\sqrt{a}.
	$ Combining the two contributions, the minimum error probability for the overlapping-region case near the critical point takes the asymptotic form
	\begin{equation}
		P_{\rm e}^{\rm min}
		\simeq
		\frac{\mathcal C_{\rm ov}(t_*)}{\sqrt N}f_{\rm ov}(N\xi)
		+
		\frac{\mathcal C_{\rm s}(t_*)}{\sqrt N}f_{\rm s}(N\xi).
	\end{equation}

\end{widetext}

\end{document}